\documentclass[aps,preprint,showkeys,onecolumn,nofootinbib]{revtex4-1}
\usepackage[utf8]{inputenc}
\usepackage{amsmath}
\usepackage{amsfonts}
\usepackage{amssymb}
\usepackage{hyperref}
\usepackage{color}
\usepackage{bbm}
\usepackage{epsfig}

\begin{document}
\author{Christian Copetti}
\email{christian.copetti@uam.es}
\author{Karl Landsteiner}
\email{karl.landsteiner@csic.es}
\affiliation{Instituto de F\'isica Te\'orica UAM/CSIC, c/Nicol\'as Cabrera 13-15, Universidad Aut\'onoma de Madrid, Cantoblanco, 28049 Madrid, Spain}

\title{Anomalous Hall viscosity at the Weyl semimetal/insulator transition}
\date{\today}

\begin{abstract}
We show that 3D Lifshitz fermions arising as the critical theory at the Weyl semimetal/insulator transition naturally develop an anomalous Hall viscosity at finite temperature. We discuss how to couple the system to non-relativistic background sources for stress-tensor and momentum currents via a form of Newton-Cartan geometry with torsion and derive the Kubo formulas for the Hall viscosities.  While the Lifshitz system that arises most naturally has scaling exponent
$z=2$ we also generalize the theory for arbitrary Lifshitz scaling $z$ and show that, in the  limit $z \to 0$, it may be given a Chern-Simons interpretation by dimensionally reducing along the anisotropic direction. The Hall viscosities are expressed in terms of zeta functions  and their temperature dependence is dictated by the scaling exponent. 
\end{abstract}
\preprint{IFT-UAM/CSIC-19-009}

\maketitle

\section{Introduction}
The presence of non-dissipative (Hall) viscosity in two dimensional materials is one of the well established hallmarks of a topological phase of matter \cite{Avron:1995fg,Hoyos:2014pba}. 
In recent years an increasing amount of interest has been devoted to understand the possibility of the emergence of such phenomena in gapless systems. In particular, two dimensional Lifshitz fermions have been shown to possess a non vanishing Hall viscosity both at finite temperature and at finite magnetic field \cite{Link:2017ora,Pena-Benitez:2018dar}, while a similar analysis in the case of the 3D fermions in magnetic field has been carried out in \cite{Offertaler:2018gwz}. However it is not clear whether such features are a universal property of critical Lifshitz theories or not. In the latter case the presence of Hall viscosity may be a definite macroscopic signature of the quantum critical point.
In parallel, a considerable amount of effort has been devoted to the formulation of effective field theory of non-relativistic quantum systems. The most prominent example of this is the use of Newton-Cartan geometry \cite{Duval:2009vt,Jensen:2014aia} to construct the effective action for quantum Hall systems \cite{Son:2013rqa,Bradlyn:2014wla}. Even in the absence of the full Galilei group, Lifshitz and anisotropic theories have been extensively investigated\cite{Hoyos:2013qna,Hoyos:2013eza,Gromov:2017gsk}.

We will study $3D$ critical Lifshitz fermions with broken time-reversal symmetry. 
Recent studies using AdS/CFT \cite{Landsteiner:2016stv} have suggested that such systems should develop a finite Hall viscosity in a thermal bath, which should be seen as characterizing the quantum critical region. Furthermore, it has been suggested that the Hall viscosity should be proportional to the mixed gauge/gravitational anomaly of the high energy fermionic theory. While such a claim is intriguing, it is hard to explain from Quantum Field theoretical considerations, since at the critical point no obvious notion of chiral symmetry is present. 
In particular we will be interested in the $z=1/2$ theory, which is expected to describe the quantum critical point of a Weyl semimetal/insulator transition \cite{RevModPhys.90.015001}\footnote{The stability fo the Lifshitz point under interactions has been shown in \cite{yang2014quantum}}. We also study the $z \to 0$ limit, which is amenable to some extent to an effective field theory treatment. The Weyl semimetal/insulator transition and the critical point can be described by starting from the UV Dirac type Lagrangian \cite{Grushin:2012mt}
\begin{equation}
\mathcal{L}= \bar{\psi} \left( i \gamma^\mu \partial_\mu -m + \gamma^\mu \gamma_5 b_\mu \right) \psi \, .
\end{equation}
This simple model is known to have two quantum phases. When $b^2+m^2 < 0$  the low energy physics is described by two Weyl nodes displaced in momentum space whereas for $b^2+m^2>0$ a gap is present\footnote{See the appendix \ref{app:wsmdetails} for further discussion}. At the critical point the system develops a quadratic energy dispersion in the $b$ direction, and its low energy physics may be described by an anisotropic two-component fermionic Hamiltonian.
This can be seen explicitly by choosing a convenient basis of gamma matrices, $\gamma^\mu =\{ \tau_3\otimes\sigma_3,i\tau_3\otimes\sigma_2, -i\tau_3\otimes\sigma_1, i \tau_2\otimes \mathbbm{1}\}$ with $\tau_i$ and $\sigma_i$ denoting two copies of the standard Pauli matrices. 
Assuming $b_\mu$ to be spacelike we chose coordinates such that it points in the $3$ direction.
The Dirac type Hamiltonian is then
\begin{equation}
    H = \begin{pmatrix}
    \sigma_\perp.k_\perp +(b+m) \sigma_3 & \sigma_3 k_3 \\
    \sigma_3 k_3 & \sigma_\perp.k_\perp +(b-m) \sigma_3
    \end{pmatrix}\,.
\end{equation}
For large $|m+b|\gg |k|$ the four component spinor $(\phi,\psi)$ can be reduced to a two component
spinor by setting $\phi = - k_3 \psi/(m+b) \psi$. In the case $|b-m|\gg |k|$ one solves instead for the spinor components $\psi$.
We note that the charge conjugation matrix in this representation is $\mathcal{C} =  i\mathbbm{1} \otimes\sigma_2$. In particular this means that charge conjugation is a symmetry of the effective two band Hamiltonian acting on $\psi$
\begin{equation}\label{eq:2bandH}
H = \sigma_\perp.p_\perp +  \sigma_3 (s p_3^2+\Delta) \,.
\end{equation}
Compared to the four band model we have rescaled momenta by setting $k_3^2/|b+m| \rightarrow p_3^2$ and $k_\perp \rightarrow p_\perp$ and wrote $\Delta = b-m)$ and $s= - \mathrm{sgn}(b+m)$.
The model is gapped for $s \Delta>0$, in a Weyl semimetal phase for 
$s \Delta<0$. At $\Delta =0$ there is a critical point with anisotropic Lifshitz scaling symmetry $p_3  \to \lambda^{1/2} p_3$ , $(\omega,p_\perp) \to \lambda (\omega,p_\perp)$. Here $s=\pm 1$ sets the direction of fusion between the chiral Weyl points. One may say that $s$ acts as a remnant of the emergent chiral symmetry of the model.  From now on we will study the critical theory at $\Delta=0$.
 
\begin{figure}[!thb]
\begin{center}
\includegraphics[scale=0.55,clip=true]{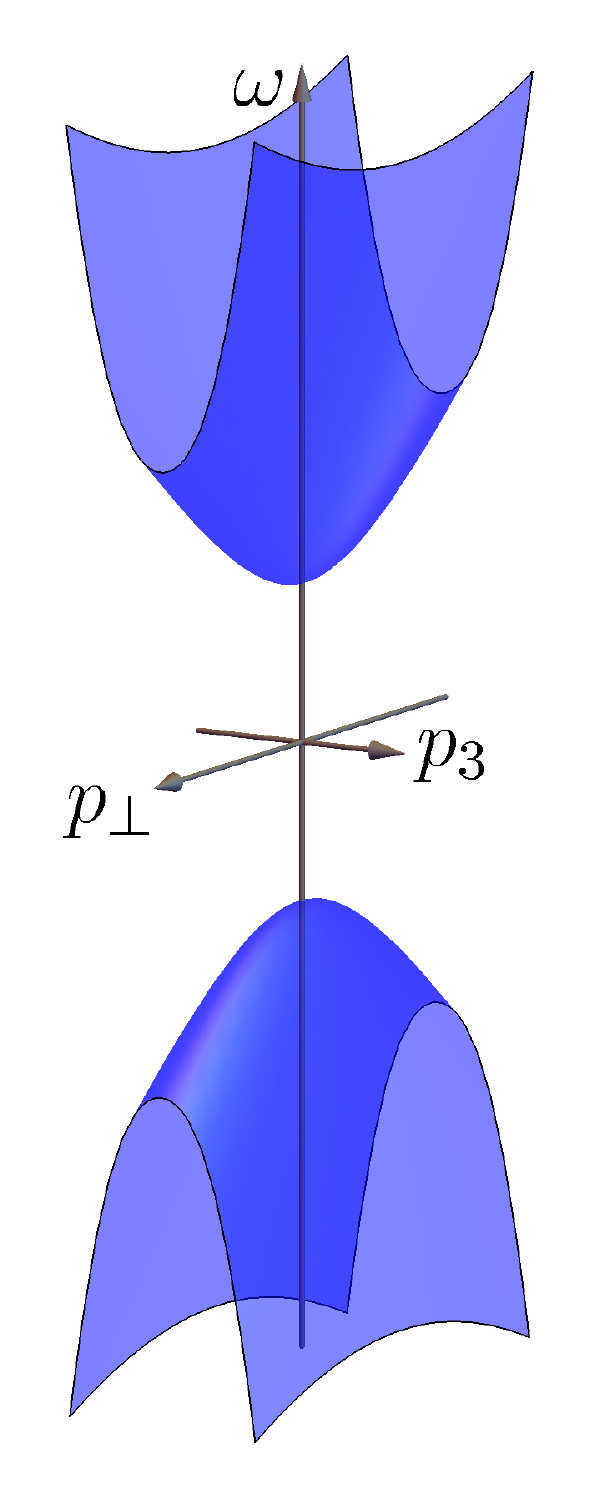}
\includegraphics[scale=0.55,clip=true]{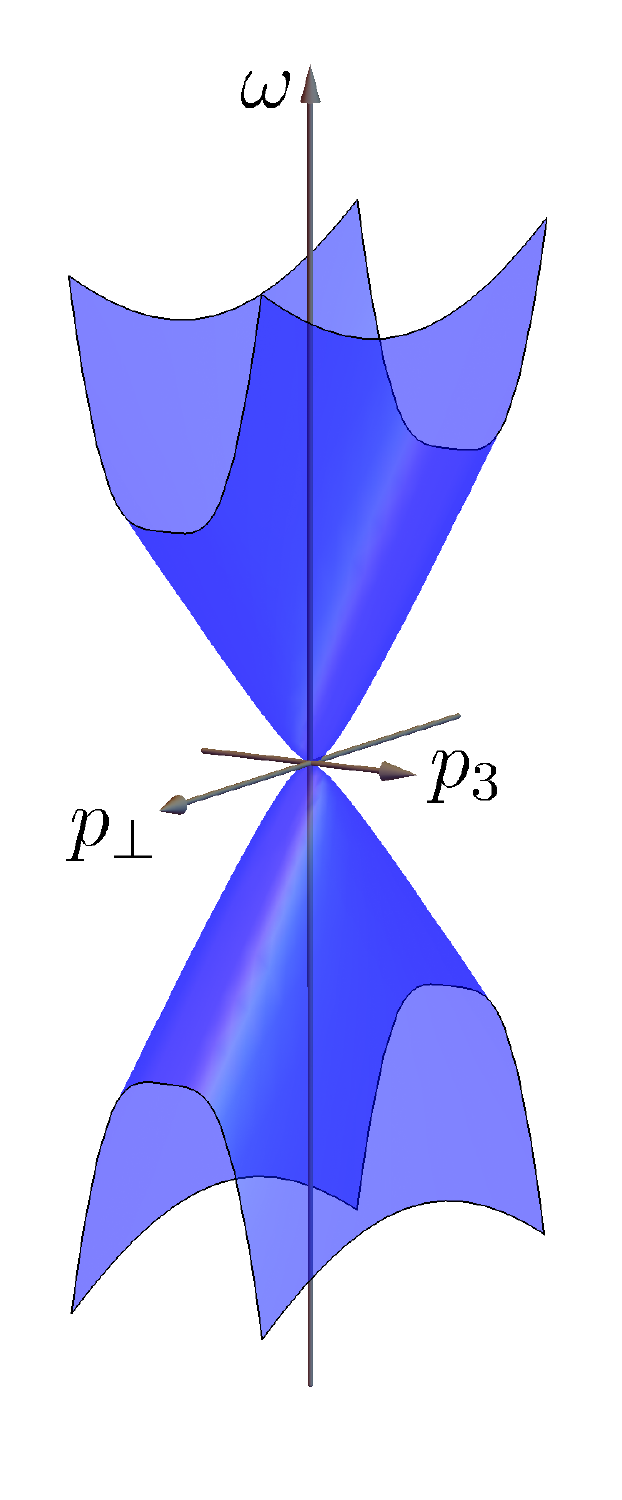}
\includegraphics[scale=0.55,clip=true]{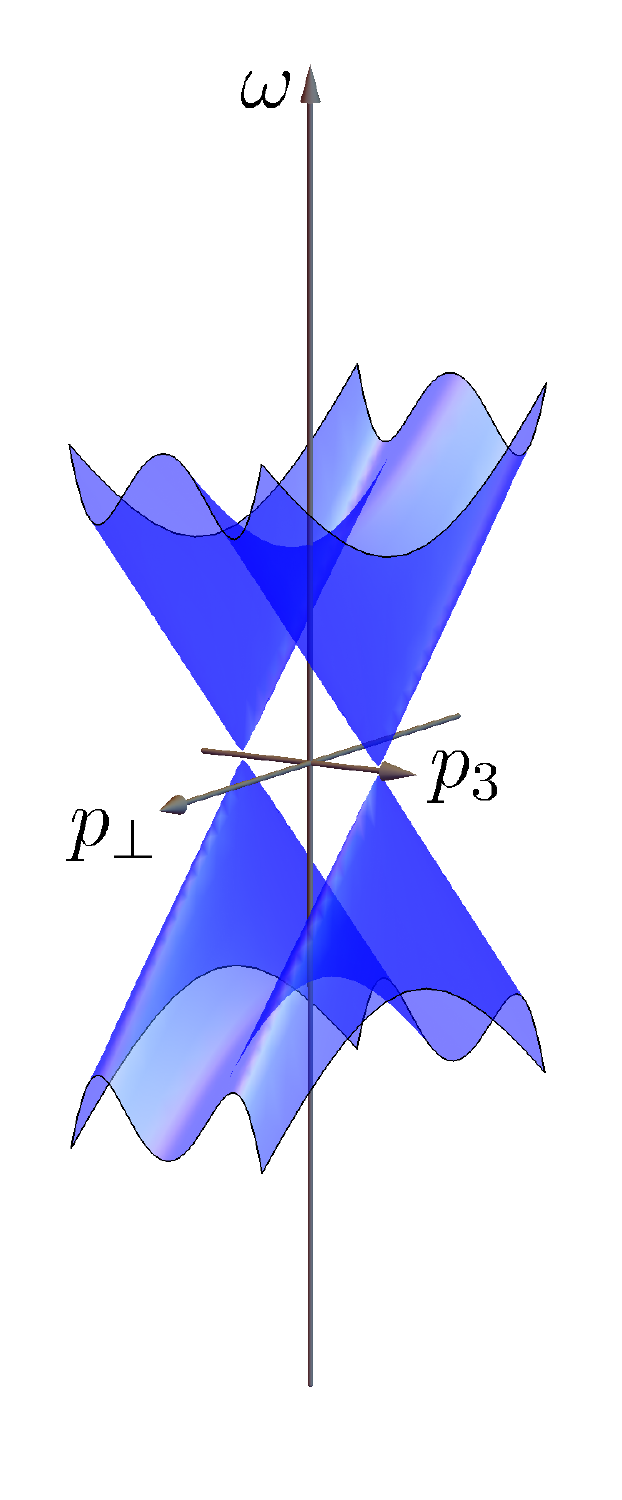}
\end{center}
\vskip -4mm 
\caption{Phases of the two-band Hamiltonian \ref{eq:2bandH}
The figure shows the dispersion relation as function of $p_\perp$ and $p_3$. The left figure is the insulating phase.
The right figure the is deep in the Weyl semimetal phase and the middle figure shows the critical point in between the two.} 
\label{fig:phases}
\end{figure}

To discuss symmetries and coupling to background fields it is slightly more natural to switch to a Lagrangian formulation. Since the rotation group is broken, we need to work with fermionic degrees of freedom transforming under the reduced rotation group $SO(1,2)$ only. These are just familiar $2+1 $ dimensional fermions $\varphi$. The appropriate $\gamma$ matrices $\gamma_A$, $A\in \{0,1,2\}$  up to unitary equivalence are taken to be $\gamma^A=(\sigma_3, -i \sigma_2, i\sigma_1)$. 
It is well known that the Lorentzian Clifford algebra in $(2+1)$ dimensions allows a Majorana representation consistent
with the already stated invariance of the Hamiltonian under charge conjugation with charge conjugation matrix $C= i\sigma_2$. The Lagrangian is
\begin{equation}\label{eq:2bandlagrangian}
{\mathcal L}= \bar{\varphi}(-p)\left[\gamma_A p^A + \mu(p) \right] \varphi(p) \, , 
\end{equation}
where $\bar \varphi = \varphi^\dagger \gamma^0$. In this language the anisotropic term will act as a momentum dependent mass $\mu(p)= s p_3^2$ whose sign is given by $s$. Time reversal flips the sign of the fermion mass in $2+1$ dimensions, which in our case amounts to $s \to - s$.
Thus a time reversal invariant system has at least two copies of the Lagrangian eq.(\ref{eq:2bandlagrangian}) with opposite
choices for $s$.
The minimal model is one of a single Majorana fermion $\chi$ obeying $C.\bar\chi^T = \chi$.

A generalization is to take the Lifshitz scaling exponent arbitrary $(\omega, p_\perp, p_3) \rightarrow (\lambda \omega,\lambda p_\perp, \lambda^z p_3)$. In this case the momentum dependent mass term takes the form $\mu(p)= s  |p_3|^{1/z}$.
Our Lifshitz differes from the one usually employed in the literature in that the anisotropic direction is a space direction and not time, as is the case for example in Galileian physics. 
We also note that in the limit $z\rightarrow 0$ the momentum in the third direction $P_3$ becomes a central element
of the Lifshitz algebra since $[D,P_3] = -z P_3$ and $P_3$ commutes with the other generators.

We will compute the Hall viscosity tensor for this class of models in the linear response regime, showing that indeed it is nonzero at finite temperature.
In order to do this, we will couple the system to a curved space-time with non-vanishing torsion that will allow us to properly define the stress generators and the Kubo formulas in the Lifshitz case\footnote{A similar approach in the 2D case was developed in \cite{PhysRevLett.107.075502,PhysRevD.88.025040}}.
The paper is organized as follows. In section \ref{sec:NCgeometry} we present the non-relativistic Newton-Cartan type geometry. In section \ref{sec:Hviscosity} we give the main steps in the Kubo formulas computation, summarize the results on Hall viscosities and comment on the simplifications happening in the  $z\to 0$ limit of \eqref{eq:2bandlagrangian} from the point of view of effective field theory. In section \ref{sec:discussion} we conclude with a few remarks and open questions for further discussion. The (many) technical details are relegated to the appendix \ref{sec:appendix}.
Throughout we use greek letters $\mu,\nu, \rho \, ...$ for spacetime indexes, lower-case latin letters $a,b,c \, ...$ for  $SO(1,3)$ tangent space indexes and upper case latin letters $A,B,C \, ...$ for the unbroken $SO(1,2)$ tangent space indexes.

\section{Coupling to curved spacetime}\label{sec:NCgeometry}
A first step in determining the properties of a system is to examine its symmetries. In particular we will be interested in the way the symmetry currents of our Lifshitz  system couple to (external) gauge fields. This allows us to derive the most general form for the conserved currents and the link to their responses to external perturbation through the Kubo formalism.

While standard relativistic systems with the full Lorentz symmetry couple to a pseudo-Riemannian geometry, this is in general not possible for their non-relativistic analogues. In this section we review the geometric structure to which a Lifshitz theory should couple and explain how it can be recovered as a limit of Newton-Cartan geometry.
As a byproduct, we will see that a curved spacetime version of \eqref{eq:2bandlagrangian} theory emerges as the lowest order derivative action which breaks $T$ symmetry with Lifshitz scaling. 

What we want to implement is a geometry which, together with the usual diffeomorphism invariance, has a preferred (covariantly constant) one-form field $l_\mu$, which will reduce to $\delta_\mu^3$ in the flat limit.

Once this one form is specified, there are various ways to approach the problem. One is to follow the standard treatment of Newton-Cartan geometry \cite{Duval:2009vt,Jensen:2014aia} and then restrict the set of geometric data to be compatible with the Lifshitz scaling symmetry. 
We will follow an ultimately equivalent prescription, commenting in the end about the connection with Newton-Cartan geometry.

\subsection{Geometry}
The starting point for us will be a spacetime metric $g_{\mu\nu}$ and a one-form field $l_\mu$ which is normalized to one $l^\mu l_\mu =1$, being $l^{\mu}= g^{\mu\nu}l_\nu$. This defines a splitting of the metric
\begin{equation}
g_{\mu\nu}= l_\mu l_\nu + h_{\mu\nu} \, ,   
\end{equation}
where $h_{\mu\nu} l^\mu=0$. 
To define the geometry we further need to define the parallel transport of tensors, which requires specifying a connection $\Gamma$ on our manifold to build a covariant derivative $\nabla$. This acts on tensors as
\begin{equation}
\nabla_\mu V^\alpha_\beta= \partial_\mu V^{\alpha}_\beta + \Gamma^\alpha_{\gamma\mu} V^\gamma_\beta - 
\Gamma^\gamma_{\beta\mu} V^\alpha_\gamma \, .
\end{equation}
We will require the metric to be covariantly constant $\nabla_\mu g_{\alpha \beta}=0$. This fixes the connection to the Levi-Civita form plus the undetermined contorsion tensor \cite{shapiro2002physical} ${K^\lambda}_{\mu\nu}=\frac{1}{2}\left({T^\lambda}_{\mu\nu} - {{T_\mu}^\lambda}_\nu -  {{T_\nu}^\lambda}_\mu\right)$, being ${T^\lambda}_{\mu\nu}$ the torsion.
\begin{equation}
{\Gamma^{\rho}}_{\mu\nu} = \frac{1}{2}g^{\rho\tau}\left(-\partial_\tau g_{\mu\nu}  +\partial_\mu g_{\nu\tau} + \partial_\nu g_{\mu\tau} \right) + {K^\rho}_{\mu\nu} \, .
\end{equation}
We will suppose that the torsion is purely of the form ${T^\lambda}_{\mu\nu}= l^\lambda T_{\mu\nu}$. Then demanding $l_\mu$ to be covariantly constant fixes this to
equation gives the conditions
\begin{equation}
T_{\mu\nu}= - \left(\partial_\mu l_\nu - \partial_\nu l_\mu \right) \, .
\end{equation}
The previous two conditions imply that $\nabla_\mu h_{\alpha\beta}=0$, which constrains 
\begin{equation}
\mathcal{L}_l h_{\mu\nu}=0 \, ,
\end{equation}
$\mathcal{L}$ being the Lie derivative. Even so, $l^\mu$ will not be a Killing vector for the metric in the presence of torsion
\begin{equation}
\mathcal{L}_l l_\mu= T_{\alpha \mu} l^\alpha \equiv G_\mu \, .
\end{equation}
After some algebra one can write the connection as \begin{equation}
{\Gamma^\rho}_{\mu\nu}= l^\rho \partial_\nu l_\mu + \frac{1}{2}h^{\rho \sigma} \left( \partial_\mu h_{\sigma \nu} + \partial_\nu h_{\sigma \mu} - \partial_\sigma h_{\mu\nu}  \right) = l^\rho \partial_\nu l_\mu + {\hat\Gamma^\rho}_{\mu\nu}[h] \, .
\end{equation}
For a bosonic system this is enough to determine completely the coupling to geometry, however, since we are dealing with fermions, we will also need vielbein fields $e_\mu^A$ which couple to the internal spin degrees of freedom of the fermionic particles. 
These are defined through the splitting $h_{\mu\nu}= e^A_\mu e^A_\nu \eta_{AB}$. They also satisfy $e_\mu^A l^\mu=0$. We will also introduce inverse vielbein fields $E^\mu_A$ defined through the orthonormality conditions
\begin{equation}
e_\mu^A E^\mu_B= \delta^A_B \,, \ \ \  l_\mu E^\mu_A =0 \, .
\end{equation}
As customary, we also introduce a spin connection ${{\omega_\mu}^A}_B$ which acts on fermionic fields and on the vielbein.
Given the connection $\Gamma$ this is uniquely determined as a function of the geometric data by demanding the vielbein to be covariantly constant
\begin{equation}
\nabla_\mu e_\nu^A = \partial_\mu e_\nu^A - \Gamma^\gamma_{\nu\mu} e_\gamma^A + {{\omega_\mu}^A}_B e_\nu^B = 0 \, ,
\end{equation}
as 
\begin{equation}
{\omega_\mu}^{AB}= -E^{\nu A} \left( \partial_\mu  e_\nu^B  -  {\hat\Gamma^\rho}_{\nu\mu}[h] e_\rho^A \right) \, .
\end{equation}
Notice that in this case the spin connection is torsion less, in form language $d e^A + \omega^A_B \wedge e^B\equiv T^A=0$.

Let us compare this construction to the one in the Newton-Cartan formalism. Let us start by defining a Newton-Cartan structure through a one form $l_\mu$ and a symmetric twice covariant tensor $h^{\mu\nu}$ whose kernel is spanned by $l_\mu$, namely $h^{\mu\nu}l_\nu =0$. 
One can further define the vector $l^\mu$ and the symmetric twice contravariant tensor $h_{\mu\nu}$ through the algebraic relations
\begin{equation}
l^\mu l_\mu = 1 \, , \ \ l^\mu h_{\mu\nu}=0 \, , \ \ h^{\mu\alpha} h_{\alpha \nu}= \delta^\mu_\nu -l^\mu l_\nu \, ,
\end{equation}
such that $h^{\mu\alpha} h_{\alpha \nu}= P^\mu_\nu$ is a projector orthogonal to both $l^\mu$ and $l_\mu$. 
The ambient metric is then defined as
\begin{equation}
g_{\mu\nu} = l_\mu l_\nu + h_{\mu\nu} \, .
\end{equation}
To define our geometric setup we further need to specify a connection to parallel transport tensors. The standard way of doing this is by demanding the original data to be covariantly constant
\begin{equation}
\nabla_\mu l_\nu = \nabla_\mu h^{\alpha \beta} = 0 \, ,
\end{equation}
with a further restriction that the torsion tensor $T^\lambda_{\alpha\beta}$ satisfies 
\begin{equation}
h_{\tau \lambda} T^\lambda_{\alpha\beta} = 0 \, ,
\end{equation}
Solving these equations fixes the connection to the same form we have found apart from an undetermined two-form $F_{\mu\nu}$
\begin{equation}
\Gamma^\mu_{\nu\rho}= l^\mu \partial_\rho l_\nu + \hat\Gamma^\mu_{\nu\rho}[h] + h^{\mu\sigma}l_{(\nu} F_{\rho)\sigma} \, ,
\end{equation}
furthermore, our data are not completely specified, indeed the Milne boosts
\begin{align}
{l'}^\mu &= l^\mu + h^{\mu\nu} \Psi_\nu \, , \\ \label{Milne}
{h'}_{\alpha \beta} &=  h_{\alpha\beta} - 2 l_{(\alpha} P^\nu_{\beta)} \Psi_\nu + l_\alpha l_\beta h^{\mu\nu} \Psi_\mu \Psi_\nu   \, .
\end{align}
leave the orthonormality relations invariant. These two pieces of data are indeed problematic for our Lifshitz effective theory, in fact, they are responsible, respectively, for the $U(1)$ particle number (for which $F_{\mu\nu}$ is interpreted as a field strength) and Galilean boots symmetries of non relativistic theories.
However we will be interested in theories that are invariant under charge conjugation, so that a real representation of the relevant degrees of freedom should exist. This suggests that we should set $F_{\mu\nu}=0$. In parallel, Lifshitz theories with $z \neq 2$ do not seem to be compatible with the Milne redefinition above, \cite{grinstein2018existence}. We should thus fix the Milne frame by some physical consideration. 
A useful way to fix $\Psi_\nu$ is to notice that, with our choice for the connection, neither $l^\mu$ nor $h_{\alpha \beta}$ are covariantly constant a quick calculation setting $F_{\mu\nu}=0$ gives \cite{Son:2013rqa}
\begin{align}
\nabla_\mu l^\nu &= \frac{1}{2} h^{\alpha \nu}\mathcal{L}_{l} h_{\alpha \mu} \, , \\
\nabla_\mu h_{\alpha\beta} &= l_{(\alpha} \mathcal{L}_l h_{\beta)\mu} \, ,
\end{align}
being $\mathcal{L}$ the Lie derivative. Let us stress that these equations are not independent, but one implies the other once the orthogonality condition $l^\mu h_{\mu\nu}=0 $ is imposed.
One then sees that our geometry corresponds to a Newton-Cartan setting in which no boots symmetry is allowed and no $U(1)$ symmetry is present either.

Of course it would be interesting to understand if generalizations are possible in order to still accommodate fermionic Lifshitz systems with $z \neq 2$, but for the present work we will not need such further generalizations.

\subsection{Ward identities}
Now that we have defined the geometric background, it is useful to briefly derive the Ward identities obeyed by the currents which couple to our set of external fields $\lbrace e_\mu^A, l_\mu, {{\omega_\mu}^A}_B , T_{\mu\nu} \rbrace$. This will bring about an important point about the nature of the independent data we will be using. In fact, as customary, regarding the connection (and in this case the non vanishing torsion) as functions of $e_\mu^A$ and $l_\mu$ bring about an improvement of the conserved currents. This may manifest itself in the linear response formulation, giving rise to different transport coefficients. In the following we will stay faithful to the Quantum Field Theory literature, in which the improved currents are used as generators for the symmetries, this is the natural choice if the spin connection is torsion less

We begin by writing down a general variation of the effective action 
\begin{equation}
\delta S= -\int \sqrt{g} \left(t^\mu_A \delta e_\mu^A + p^\mu \delta l_\mu + {S^\mu}_{AB} \delta {\omega_\mu}^{AB} + \Omega^{\mu\nu} \delta T_{\mu\nu} \right) \, .
\end{equation} 
Here ${t^\mu}_A$ is the unimproved stress tensor, ${S^\mu}_{AB}$ the spin current and $\pi^\mu$ the anisotropic momentum current. The inverse vielbein and the vector $l^\mu$ are treated as dependent objects, whose variation is re expressed by using
\begin{align}\label{varfields1} 
\delta E^\nu_B &= - E^\mu_B E^\nu_A \delta e^A_\mu  -l^\nu E^\mu_B \delta l_\mu\,\\ 
\delta l^\nu &= - l^\mu E^\nu_A \delta e^A_\mu - l^\mu l^\nu\delta l_\nu .
\end{align}
The Ward identities follow from the local invariance of the action under diffeomorphism and tangent space rotations on the independent fields, these read
\begin{align}
\delta_\xi l_\mu &= \nabla_\mu (\xi^\nu l_\nu) - T_{\nu \mu} \xi^\nu \, , \\
\delta_\xi e_\mu^A &= \nabla_\mu (\xi^\nu e_\nu^A) - \xi^\lambda {\omega_\lambda}^{AB} e_\mu^B\, , \label{vielbeindiffeo}
\end{align}
for the diffeomorphism variation and
\begin{align}
\delta_\Omega l_\mu &=0 \, \\
\delta_\Omega e_\mu^A &= {\Omega^A}_B e_\mu^B \, .
\end{align}
for tangent space rotations. 
The last term in \eqref{vielbeindiffeo} is not covariant under tangent space transformations, as is the case for connections. However we may combine it together with a Lorentz variation with $\Omega^{AB}_\xi = \xi^\lambda {\omega_\lambda}^{AB}$ to cancel it. We will use such "covariantized" variation in what follows.

In view of the application of the Kubo formalism, we will find useful to saturate the spacetime indexes of the objects by contracting either with the vielbein or the vector $l^\mu$ in order to better distinguish Lorentz invariant objects. Thus we will often use splittings of the form $V^\mu = l^\mu v + E^\mu_A v^A$. Splitting the diffeomorphism generator $\xi^\mu = \theta l^\mu + E^\mu_A \xi^A$ gives for the covariant diffeomorphism variation
\begin{align}
\delta_\theta l_\mu &= \partial_\mu \theta - \theta G_\mu \, , \\
\delta_\theta e_\mu^A &= 0 \, ,\\
\delta_\xi l_\mu &=  -T_{A \mu} \xi^A \, , \\
\delta_\xi e_\mu^A &= \nabla_\mu \xi^A \, .
\end{align}
The variation of the spin connection is recovered by using the identity
\begin{equation}
\delta {\omega_\mu}^{AB} = -\frac{1}{2} \left( E^{\nu A} \nabla_\mu \delta e_\nu^B + E^{\nu B} \nabla_\nu \delta e_\mu^A - E^{\nu A} E^{\mu B} e_{\mu C} \nabla_\nu \delta e_{\rho}^C \right) - (A \leftrightarrow B) \, .
\end{equation}
This, together with the explicit dependence of $T_{\mu\nu}$ on $l_\mu$ gives defines the improved currents
\begin{equation}
\begin{aligned}
{\tau^\mu}_A &= {t^\mu}_A  + \frac{1}{2} l^\mu (\nabla^B-G^B) \sigma_{BA} \\
&+ \frac{1}{2} \left[E^{\mu B} (\nabla^C-G^C) \left(s_{CBA} +s_{BAC} - s_{ABC} \right) +\nabla_l \sigma_{BA} \right] \, ,
\end{aligned}
\end{equation}
and
\begin{equation}
\pi^\mu = p^\mu - \left( \nabla_\nu - G_\nu \right) \Omega^{\nu\mu} \, ,    
\end{equation}
where $\nabla_l \equiv l^\mu \nabla_\mu$, whereas $s_{ABC}$ and $\sigma_{AB}$ are defined through the splitting of the spin connection by \begin{equation}
{S^\mu}_{AB}= E^{\mu C} s_{CAB} + l^\mu \sigma_{AB} \, .
\end{equation}
To derive these formulas one needs the identity 
\begin{equation}
\frac{1}{\sqrt{g}}\partial_\mu \sqrt{g}= \Gamma^\nu_{\mu_\nu}  = \Gamma^\nu_{\nu_\mu} + G_\mu \, ,
\end{equation}
to integrate by parts in our torsionful geometry. Plugging in the variations of the independent fields  we get the diffeomorphism and Lorentz Ward identities
\begin{align}
\left(\nabla_\mu -G_\mu\right) {\tau^\mu}_A &= T_{A\mu} \pi^\mu \, , \\
\left(\nabla_\mu -2 G_\mu\right)\pi^\mu &= 0 \, , \\
e_{\mu [A} {\tau^\mu}_{B]} &= 0 \, .
\end{align}
which can be recast by further saturating the contracted spacetime indexes through
\begin{align}
{\tau^\mu}_A &= E^{\mu B} \tau_{BA} + l^\mu \Sigma_A \, , \\
\pi^\mu &= E^{\mu A} \pi_A + l^\mu \pi \, ,
\end{align}
so that the diffeomorphism and Lorentz Ward identities read
\begin{align}
\left( \nabla_A - G_A \right) \tau^{AB} + \nabla_l \Sigma^B &= {T^B}_A \pi^A + G^B \pi \, , \\
\left(\nabla_A -2 G_A\right)\pi^A + \nabla_l \pi &= 0 \, , \\
\tau_{[AB]} &= 0 \, .
\end{align}

Furthermore, since of theory is also Lifshitz invariant, one may introduce the following transformation rule under Weyl rescalings
\begin{align}
\delta_\sigma l_\mu &= z \sigma l_\mu \, , \\
\delta_\sigma e_\mu^A &= \sigma e_\mu^A
\end{align}
which give rise to the Lifshitz Ward identity
\begin{equation}
{\tau^A}_A + z \pi =0 \, .  
\end{equation}

The improved stress tensor $\tau_{AB}$, anisotropic momentum $\pi_A$ and anisotropic strain $\Sigma_A$ will be the quantities used in the linear response formulation.

\section{Lifshitz hydrodynamics and Kubo formulas for anisotropic Hall viscosity}\label{sec:hydro}
Now we develop a linear response formalism for the strain deformations, that is for response to changes in the external vielbein $e_\mu^A$ and $l_\mu$. This will give us a clear definition of the relevant Kubo formulae, together with the necessary contact (seagull) terms that may arise during the computation. 

In doing this we also make contact with the hydrodynamic expansion for a fluid in a Lifshitz spacetime, in which case the viscosity tensor is defined through the response of the stress tensor to a velocity gradient.
Since the systems we are going to study have a spacelike, rather than timelike, vector field dictating the anisotropic direction, we will end up with a system quite different from previous studies \cite{Hoyos:2013qna,Hoyos:2013eza} and from Galilean hydrodynamics.
The reason is that we cannot identify our vector field $l_\mu$ with the velocity field of the long distance hydrodynamic description as it is customarily done. Rather the two have to be introduced separately and with a reduced tangent space bundle in order to consistently couple a Lifshitz spacetime. 
In the end we will see that the link between viscosity (that is response to velocity gradients), from the perspective of an external relativistic observer, and time variation of the vielbein, is not accurate for gradients of the $l_\mu$ components of the velocity field. Instead such gradients provide no geometric response which is however encoded in the torsional response of the anisotropic momentum current $\pi_A$. We provide physical intuition behind this picture at the end of the Section.

As always one should start the hydrodynamic formulation by introducing a velocity vector vield. In a fully relativistic theory this may though of as a tangent vector $u^a$ normalized to $u^a u_a =-1$, this is related by a local Lorentz boost to the rest-frame field $u^a=(1,\vec{0)}$. The hydrodynamic equations then follow by substituting in the Ward identities for the conserved current the most general expansion for their one point functions in terms of gradients of the velocity vector and (possibly) external gauge fields.

In our case however, the boots symmetry is restricted, so that, if we define a velocity field
\begin{equation}
u^\mu = \theta l^\mu + v^A E^\mu_A \, ,
\end{equation}
only the latter part of the above expression may be brought in a canonical form $v^A=(\rm{v},\vec{0})$ via a local Lorentz boots.

Thus in our case the anisotropic velocity $\theta$ should be viewed as an intrinsic property of the flow and it will be instructive to divide such flows in two parts, depending on whether or not $\theta=0$.

Another, probably more intuitive interpretation is as follows. In a flat geometry with $e^A_{\mu} = \delta^A_{\mu}$ and 
$l_\mu = \delta^3_{\mu}$ we have the conservation equation $\partial_A \pi^A + \partial_3 \pi =0$. This shows that $\int d^3x \pi^0$ is a conserved charge and we can define a grand canonical ensemble with chemical
potential conjugate to this charge. In fact this charge is nothing but the momentum in the $3$ direction. This chemical
potential should be identified with the parameter $\theta$  in the same way as fluid velocity $v^A$ is the chemical potential
for the other momentum components. In this interpretation the we define the restframe as $v^A=(1,0,0)$ and $\theta=0$.

Let us start by considering the case $\theta=0$. We will work in a derivative expansion around the rest-frame $v^A=(1,0,0)$ and in metric perturbations around the "flat" geometry $e^A_{\mu} = \delta^A_{\mu}$,  
$l_\mu = \delta^3_{\mu}$.

To first order in derivatives, we of course need to take into account the covariant derivative of the velocity field $\nabla_\mu v^A$. However, at the same order in derivatives we should also keep track of another independent data in our chosen geometry. This is the background torsion
\begin{equation}
T_{\mu\nu}= -\left(\partial_\mu l_\nu - \partial_\nu l_\mu \right) \, .
\end{equation}
These data are now to be projected such that they are orthogonal to the velocity field $v^A$, through the projector
\begin{equation}
P^A_B = \delta^A_B + v^A v_B \, ,
\end{equation}
which we often leave implicit to avoid cluttering of notation.
Furthermore spacetime indexes, when present, will be saturated using the geometric data $e_\mu^A$, $l^\mu$ and then projected.
This gives the following set of data
\begin{equation}
\nabla_\mu v_A = l_\mu \nabla_l v_A + e_\mu^B \left(\hat{\sigma}_{AB} +\eta_{AB} \Theta + \epsilon_{AB} \omega \right) \, ,
\end{equation}
having defined the shear $\hat{\sigma}_{AB} = \nabla_{(A} v_{B)} - \frac{1}{2} \eta_{AB} \nabla_C v^C$, the expansion $\Theta= \nabla_C v^C$ and the vorticity $\omega= \epsilon^{ABC} v_A \nabla_B v_C$, with $\epsilon_{AB}= \epsilon_{ABC} v^C$. 
In much the same way, the torsion tensor also has en electric-magnetic decomposition through
\begin{equation}
T_{\mu\nu} = 2(l_{[\mu} e_{\nu]}^A G_A + e_{[\mu}^A e_{\nu]}^B \left( \zeta_{[B} v_{A]} + \epsilon_{AB} m  \right)) \, ,
\end{equation}
here $\zeta_A$ and $m$ are the analogues of electric and magnetic field for three dimensional electrodynamics, with torsion playing the role of field strength From an ambient metric point of view the magnetic component is somewhat analogous to a gravitomagnetic field. Note that contrary to the usual case here this
"gravitomagnetic" field is a covariant tensor and can appear independently in the response. In this sense it seems related
to a response pattern that is familiar from the chiral vortical effect \cite{Amado:2011zx}. For now we defer study of anomalous transport patters analogous to chiral vortical (and chiral magnetic) effects in the Lifshitz model to future investigation, although we present some partial results in the discussion section. and concentrate on viscosity type of responses.

At this point we would be ready to develop the most general hydro response for our Lifshitz-type theories. However for the present work let us focus on the non-dissipative, time-dependent responses in the strain tensor and the anisotropic momentum current.

First let us briefly remind the reader the basic definitions of the viscosity tensor. In isotropic theories, this is defined as the response of the strain to gradients of the velocity fields, that is
\begin{equation}
\langle \tau^{\mu\nu} \rangle = \eta^{\mu\nu\rho\sigma} \nabla_\rho u_\sigma + O(\nabla^2) \, ,
\end{equation}
due to the symmetry of the strain tensor it satisfies $\eta^{\mu\nu\rho\sigma}=\eta^{\nu\mu\rho\sigma}=\eta^{\mu\nu\sigma\rho}$ where the last equality follows from the fact that the viscosity may be computed as a two point function of strain tensors. The viscosity tensor, furthermore, may be divided in a dissipative and non-dissipative (Hall) part according to the symmetry of the two couples of indexes
\begin{equation}
\eta_D^{\mu\nu\rho\sigma}=\eta_D^{\rho\sigma\mu\nu} \, , \ \ \ \eta_H^{\mu\nu\rho\sigma}=\eta_H^{\rho\sigma\mu\nu} \, .
\end{equation} 
The dissipative part of the viscosity may be further decomposed in symmetric traceless (shear) and trace-part (bulk) viscosities, while the Hall viscosity requires the introduction of a dimension-dependent tensor. In 2D this is given by the projector
\begin{equation}
P^H_{\mu\nu\rho\sigma}=  \frac{1}{4}\left( h_{\mu\rho} \epsilon_{\nu\sigma} + h_{\nu\rho} \epsilon_{\mu\sigma} + h_{\mu\sigma} \epsilon_{\nu\rho} + h_{\nu\sigma} \epsilon_{\mu\rho} \right) \, ,
\end{equation}
with $\epsilon_{\mu\nu}= \epsilon_{\mu\nu\rho} u^\rho$ and $h_{\mu\nu}=g_{\mu\nu}+ u_\mu u_\nu$. It is clear that, in 3+1 dimensions, one then needs also the presence of an additional vector field, say $b_\mu$, orthogonal to the velocity field to mimic this construction, now using $\tilde{\epsilon}^{\mu\nu}= \epsilon^{\mu\nu\rho\sigma}b_\rho u_\sigma$ to construct the projector
\begin{equation}
    \tilde{P}^H_{\mu\nu\rho\sigma}=  \frac{1}{4}\left( h_{\mu\rho} \tilde{\epsilon}_{\nu\sigma} + h_{\nu\rho} \tilde{\epsilon}_{\mu\sigma} + h_{\mu\sigma} \tilde{\epsilon}_{\nu\rho} + h_{\nu\sigma} \tilde{\epsilon}_{\mu\rho} \right) \, .
\end{equation}
This is not however the only tensor structure with the required properties, in fact
\begin{equation}
\Pi^{(1)}_{\mu\nu\rho\sigma}= b_\mu b_\rho \tilde{\epsilon}_{\nu\sigma} \, , \ \ \Pi^{(2)}_{\mu\nu\rho\sigma}= \Pi^{(1)}_{\nu \mu \sigma \rho} \, , \ \ \Pi^{(3)}_{\mu\nu\rho\sigma}=\Pi^{(1)}_{\mu\nu\sigma\rho} +  \Pi^{(1)}_{\nu\mu\rho\sigma} \, ,
\end{equation}
satisfy the required conditions. Thus one expects four independent Hall viscosity components to be present. In our formulation the fixed vector $b_\mu$ is substituted by $l_\mu$  and the corresponding indexes are automatically saturated, thus we remain with only two projectors $P_{ABCD}=\epsilon^{(A(C}\eta^{B)D)}$ and $\epsilon_{AB}=\epsilon_{ABC} v^C$, while we have to explicitly distinguish the operators $\tau_{AB}$, $\Sigma_A$, $\pi_A$ due to the lack of Lorentz invariance.

The most general expansion for the Hall coefficients then reads
\begin{align}
\langle \tau^{AB}  \rangle &= \eta_\tau^{ABCD} \hat{\sigma}_{CD}  \, \\
\langle \pi^A \rangle &= \eta^\pi \epsilon^{AB} \zeta_B + \eta^{\pi\Sigma} \epsilon^{AB} \nabla_l v_B \, \\
\langle \Sigma^A \rangle &= \eta^\Sigma \epsilon^{AB} \nabla_l v_B + \eta^{\pi\Sigma} \epsilon^{AB} \zeta_B  \, \\
\end{align} 
being $\eta_\tau^{ABCD}= \eta_\tau P^{ABCD}$. Notice that at this stage we have not included gradients of $\theta$, since they don't directly respond to geometry.

To derive Kubo formulae for the above coefficients we expand the above to first order in the external geometric data, setting $v^A=(1,\vec{0})$ to its rest frame value.
This gives, by using
\begin{equation}
\nabla_\mu v^A \sim  \partial_t e_\mu^A \, \ \ \zeta_A \sim E^\mu_A \partial_t l_\mu \, \, ,
\end{equation}
being $\partial_t= v^A \partial_A$ the time derivative.  This gives
 \begin{align}
\langle \tau^{AB}  \rangle &= \eta_\tau^{ABCD}E^\mu_C \partial_t e_{\mu D}  \, \\
\langle \pi^A \rangle &= \eta^\pi \epsilon^{AB} E^\mu_B \partial_t l_\mu + \eta^{\pi\Sigma} \epsilon^{AB} l^\mu \partial_t e_{\mu B} \, \\
\langle \Sigma^A \rangle &= \eta^\Sigma \epsilon^{AB}l^\mu \partial_t e_{\mu B} + \eta^{\pi\Sigma} \epsilon^{AB} E^\mu_b \partial_t l_\mu  \, .
\end{align} 
This leads, upon functional differentiation with respect to $e_\mu^A$, $l_\mu$, to the Kubo formulae:
\begin{align}
\eta^{\tau} &= \lim_{\omega \to 0} \frac{-i}{\omega} P_H^{ABCD}\left( G^{\tau\tau}_{ABCD}(\omega,0) + C_{ABCD}(\omega,0) \right) \\
\eta^\pi &= \lim_{\omega \to 0} \frac{-i}{\omega} \epsilon^{AB} G^{\pi\pi}_{AB}(\omega,0)  \\
\eta^\Sigma &= \lim_{\omega \to 0} \frac{-i}{\omega} \epsilon^{AB} G^{\Sigma\Sigma}_{AB}(\omega,0)  \\
\eta^{\pi\Sigma} &= \lim_{\omega \to 0} \frac{-i}{\omega} \epsilon^{AB} \left(G^{\Sigma\pi}_{AB}(\omega,0) + C_{AB}(\omega,0)\right) \, ,
\end{align}
where we have defined the retarded Green's function
\begin{equation}
G^{UV}(\omega,\vec{k}) = \int d^4 x e^{i(\omega t - \vec{k}\cdot \vec{x})} \rm{tr}\left(\rho_\beta [U(\vec{x},t), V(0,0)] \right)\theta(t) \, ,    
\end{equation}
and $C_{ABCD}$, $C_{AB}$ stand for contact terms which arise due to the explicit connection dependence of the strain generators, for our specific model they are computed in Appendix \ref{app:seagull}.

We will use such formulas in the next section to compute the odd viscosities, however let us stop for a moment to examine what we have found so far.
First notice that, contrary to the anisotropic case, there are four independent coefficients which give non-dissipative frequency dependent transport
\begin{equation}
 \eta^\tau \, , \ \ \eta^\pi \, , \ \ \eta^\Sigma \, , \ \ \eta^{\pi\Sigma} \, . \label{viscosities}
\end{equation}
In 3D such coefficients can arise only because we have broken the full rotational symmetry (which would not allow us to use the tensors $P^{ABCD}$ and $\epsilon^{AB}$).
Furthermore, under time reversal all of the above coefficients have to be odd in order to be non-vanishing Thus the microscopic theory supporting them should break such discrete symmetry.

Second, in our formulation of hydrodynamics we have not gauge the Lifshitz scaling symmetry. Imposing it on the viscosities \eqref{viscosities} through the Weyl scalings $l_\mu \to e^{-z \Omega} l_\mu$, $e_\mu^A \to e^{-\Omega} e_\mu^A$ which inverse scalings for the quantities with upper spacetime indexes
This gives the following Lifshitz scaling dimensions for the viscosities
\begin{equation}
[\eta^\tau]_L = 2 + z \, , \ \ [\eta^\pi]_L= 3z \, , \ \ [\eta^\Sigma]_L= 4-z \, , \ \ [\eta^{\pi\Sigma}]_L = 2 + z \, , 
\end{equation}
thus non-vanishing Hall viscosities need the state in which our theory is in to break the scaling symmetry. For charged particles this can be done for example by introducing a magnetic field. In our case, since we will deal with Majorana fermions, the breaking will be due to finite temperature.

Third, it should be stressed that, from the point of view of the Lifshitz theory, only $\eta^\tau$ and $\eta^\Sigma$ can be interpreted as viscosities, since they are explicitly related to gradients of the velocity field $v^A$. The coefficients $\eta^\pi$ and, in part $\eta^{\pi\Sigma}$, instead, are related to the torsional response, which is more akin to an electric conductivity.
However a moment of thought shows that the response to electric torsion need to go together with the gradients of $\theta$ in the hydrodynamic expansion. There are two ways to justify this dual description.
First one can think that, in the laboratory which as access to the full UV system, one may actually perform an $SO(1,3)$ frame redefinition, in particular, at the linearized level
\begin{equation}
l_\mu \to l_\mu'= l_\mu + \xi_A e_\mu^A \, ,    
\end{equation}
suppose we start with a geometry with vanishing $\theta$ but non-vanishing torsion. Then the transformation above sends us to a geometry with torsion
\begin{equation}
T'_{\mu\nu} =T_{\mu\nu} + 2 e_{[\mu}^A \partial_{\nu]} \xi_A + O(e^3)  \, ,    
\end{equation}
and $\theta$ component
\begin{equation}
\theta= u_A \xi^A \, ,    
\end{equation}
we may thus choose $\xi^A$ to make the torsion vanish, ending up with a nontrivial velocity gradient and vice-versa. 

Another, perhaps clearer way to see this, is to remember that in a boosted frame the hydrodynamic ensemble is constructed by coupling the conserved momentum charges $P^\mu$ to the fluid velocity $u^mu$. In particular we may define 
\begin{equation}
\tau^{\mu\nu}= \tau^{A B} E^\mu_A E^\nu_B + \pi^\mu l^\nu + l^\mu \Sigma^\nu + l^\mu l^\nu \pi\, ,
\end{equation}
putting all of the conserved currents in the same multiplet, the conserved momentum is obtained by integrating the conserved charge $T_{\mu\nu} u^\nu$ on a spatial surface of the foliation generated by $u^\mu$. Finally, ensemble coupling reads
\begin{equation}
u^\mu P_\mu = v^A \int \tau_{A t} + \theta \int \pi_t \, ,
\end{equation} 
so that $\theta$ behaves as a chemical potential for the anisotropic translation. Recall that, in the electromagnetic case, the chemical potential appears in conjunction to the electric field so that together they form the Lie derivative of the vector potential along the velocity flow (provided we choose a gauge such that $A_t =\mu$.
Then, at fixed temperature $\mathcal{L}_u A= \nabla \mu - E$ as expected.
In our case the role of the connection is played by $l_\mu$ (more precisely by $\delta l_\mu = l_\mu - \delta_\mu^3$) , its Lie derivative reads
\begin{equation}
\mathcal{L}_u l_\mu = \nabla_\mu \theta + \theta G_\mu -\zeta_A 
\end{equation}
the right hand side of this equation should be seen as the expression of "chemical equilibrium" for anisotropic translations. Notice that, in this way, the viscosity coefficients that enter though the response to the gradient of $\theta$, will also be expressible as conductivities for the electric part of the torsion.

With this understanding we can now give the physical interpretation of the different Hall viscositites in a flat background,
$e^A_\mu = \delta^A_\mu$, $l_\mu = e^z_\mu$, $A\in \{t,x,y\}$ and $\mu\in\{0,1,2,3\}$.
To this end we first note that $\tau^{AB}$ contains the energy density and the pressures in the diagonal. Its off-diagonal
entries can be interpreted either as the $x,y$ components of the energy current or the momentum densities in $x,y$ direction.
Entries with two spatial indexes are components of the strain tensor. 
The momentum density in the $z$ direction is $\Pi^t$,  $\pi$ is the $zz$ component of the
pressure. The current $\Sigma^t$ is the energy current in the $z$-direction, etc.

The first viscosity component $\eta^\tau$ is the analogue of the well known two-dimensional Hall viscosity, its is activated
if the flow and gradients are all orthogonal to $l_\mu$. If the flow is in the $x$-plane but has a gradient in the $z$-direction, $\eta^{\pi\Sigma}$ and $\eta^{\Sigma}$ describe the generation of the strain components $\Pi^y$ and $\Sigma^y$. 
If the flow is in the $z$ direction and has a gradient in the $x$ direction $\eta^{\pi\Sigma}$ and $\eta^{\pi}$ describe
the generation of $\Pi^y$ and $\Sigma^y$. We note that these last viscosities are chiral in the sense that they involve
all three directions $x,y,z$ and have a definite handedness that is determined by the parameter $s$ in the microscopic
Lagrangian.

\section{Hall viscosity of Lifshitz fermions}\label{sec:Hviscosity}
Let us now come to the question of determining whether or not the viscosities \eqref{viscosities}, even though allowed by the symmetries of the problem, are nonzero for a quantum critical theory such as Lifshitz fermions.

Also, even if it is non-vanishing, it is interesting to determine if such coefficients contain universal informations about the nature of the critical point and will not, in general, be non-vanishing for whatever T-breaking anisotropic theory we may cook up.

We will explicitly compute the value of the coefficients \eqref{viscosities} below, while at the end of the section we give a partial answer to the question of universality, at least in a particular limit.
To this end, we consider the following effective description of the Lifshitz system, which is the minimal model compatible with anisotropic Lifshitz scaling which breaks time reversal but preserves charge conjugation and parity.

\subsection{The microscopic model}

The action in the curved geometry of section \ref{sec:NCgeometry} reads
\begin{equation}
S_z = \int_\mathcal{M} \sqrt{-g} \left( \bar{\chi} i \gamma^A E_A^\mu \overset{\leftrightarrow}{\nabla}_\mu \chi + s \chi^T M(\nabla_l)^{1/2z} C^{-1} \chi \right) \, ,
\end{equation}
being $\gamma^A$ a Majorana representation of the three dimensional Clifford algebra $Cl(1,2)$, $M(\nabla_l)= \overleftarrow{\nabla}_l \overrightarrow{\nabla}_l$, $s=\pm$ the T-odd parameter. The covariant derivative acts on fermions through $\nabla_\mu \chi = \partial_\mu \chi + {\omega_\mu}^{AB} \gamma_{AB} \chi $ being $\gamma_{AB}= \frac{1}{4}[\gamma_A, \gamma_B]$ the Lorentz generators.

 Notice that, strictly speaking, the Lagrangian is local only when $z= 1/2n$. in particular $z=1/2$ represents the critical point of the Weyl semimetal insulator transition and and $z=1/2n$ can be adiabatically reached from this by tuning infrared irrelevant couplings, see Appendix \ref{app:wsmdetails}.
To work in a unified way with Majoranas is expedient to introduce the matrices $\beta^A= C^{-1} \gamma^A$ which may be represented as $\beta_0= -1$, $\beta_1= -\sigma_x$, $\beta_2 = \sigma_z$. For $A$ a spatial index these fulfill $\lbrace \beta_A, C^{-1}\rbrace = 0$ , $[\beta_1,\beta_2]= 2 C^{-1} $.

Notice also that $M(\nabla_l)$ is a positive operator, which can be seen as a mass term for 2D Majorana fermions. From this perspective the sign of $s$ is the sign of the mass of the fermionic excitations.

We will be interested in defining a strain tensor and an anisotropic momentum current for the theory in question. First one can explicitly compute the unimproved currents $t^\mu_A$, $p^\mu$ using \eqref{varfields1} to be
\begin{align}
t^\mu_A &= i E^\mu_B \chi^T \beta \overset{\leftrightarrow}{\nabla} \chi  + \frac{s}{2z} l^\mu \chi^T \left[ \overleftarrow{\nabla}_l M(\nabla_l)^{1/2z-1}  \overrightarrow{\nabla}_A +\overleftarrow{\nabla}_A M^{1/2z-1}(\nabla_l)  \overrightarrow{\nabla}_l \right]C^{-1} \chi \, , \label{unimprovedstrain}  \\
p^\mu & = i E^\mu_A \chi^T \beta^A \overset{\leftrightarrow}{\nabla}_l \chi + \frac{s}{z} l^\mu \chi^T M(\nabla_l)^{1/2z} C^{-1} \chi \, .
\end{align}
The spin current is given by ${S_\mu}^{AB}= e_{\mu C} s^{CAB} + l_\mu \sigma^{AB}$ with
\begin{align}
s_{CAB} &= i \bar{\chi} \left(\gamma_C \gamma_{AB} + \gamma_{AB} \gamma_C \right)\chi \, , \\
\sigma_{AB} &= -\frac{s}{2z} \chi^T M(\nabla_l)^{1/2z-1} \left( \overleftarrow{\nabla}_l \gamma_{AB} C^{-1} - \overrightarrow{\nabla}_l C^{-1} \gamma_{AB} \right) \chi \, ,
\end{align}
while the torsion coupling $\Omega^{\mu\nu}$ vanishes identically.

The improvement procedure has been explained in \ref{sec:NCgeometry} and can be carried out in a straightforward way. To this end notice that the spin current entering in $\tau_{AB}$ has the same structure as the isotropic free fermion one, plus contributions from $\sigma_{AB}$. Since these are by nature antisymmetric but $\tau_{[AB]}=0$ by the Lorentz Ward identity the will cancel on shell against contributions coming from the covariant derivative acting on $s_{CAB}$. The final result will be equal to the one obtained for the isotropic fermion. On the other hand the momentum currents receives no further contribution.
We thus have
\begin{align}
\tau_{AB} &= i \chi^T \beta_{(A} \overset{\leftrightarrow}{\nabla}_{B)}\chi  \, ,  \ \ \pi_A = i \chi^T \beta^A \overset{\leftrightarrow}{\nabla}_l \chi  \, , \\
\Sigma_A &= \frac{s}{2z} l^\mu \chi^T \left[\overleftarrow{\nabla}_l M(\nabla_l)^{1/2z-1}  \overrightarrow{\nabla}_A +\overleftarrow{\nabla}_A M^{1/2z-1}(\nabla_l)  \overrightarrow{\nabla}_l \right]C^{-1} \chi + \frac{1}{2} \nabla^B \sigma_{BA} \, . 
\end{align}
Notice that in the above the order of the covariant derivatives matters, since, in our geometry
\begin{equation}
[\nabla_\mu, \nabla_\nu]= 2 \partial_{[\mu} l_{\nu]} \nabla_l + {R_{\mu\nu}}^{AB} \gamma_{AB} \, ,
\end{equation}
when acting on fermions.

Even though these expressions look complicated, they simplify considerably in momentum space and we will be able to analytically extract the viscosities.

In order to compute them, we follow the standard technique for computing retarded Green's functions from analytic continuation of Euclidean ones. For this we analytically continue the Majorana fermions to Euclidean signature\cite{vanNieuwenhuizen:1996tv}.

The Euclidean correlates are then given by the following (imaginary time) Feynman diagrams
\begin{widetext}
\begin{align}
G^{\pi\pi}_{AB}(\omega)&= \frac{1}{\beta}\sum_n \int \frac{d^2 k dk_3}{(2\pi)^3} \rm{tr} \left[S(k, \omega_n) \beta_A S(k,\omega + \omega_n) \beta_B k_3^2 \right] \, \\
G^{\tau\tau}_{ABCD}(\omega) &=\frac{1}{\beta} \sum_n \int \frac{d^2 k dk_3}{(2\pi)^3} \rm{tr} \left[S(k, \omega_n) \beta_{(A} S(k,\omega + \omega_n) \beta_{(C} k_{B)} k_{D)} \right] \, , \\ 
\ C_{ABCD}(\omega) &= -\frac{\delta_{AC}}{16}\frac{1}{\beta}\sum_n \int \frac{d^2 k dk_3}{(2\pi)^3} \rm{tr} \left[\beta_{[B} \beta_{D]} \omega \beta_0 S(k, \omega_n)\right] + A \leftrightarrow B \, , C \leftrightarrow D  \, , 
\end{align}
\begin{equation}
\begin{aligned}
G^{\Sigma \pi}_{AB}(\omega) &= \frac{1}{\beta}\sum_n \int \frac{d^2 k dk_3}{(2\pi)^3} \rm{tr} \left[S(k, \omega_n)C^{-1} k_A \frac{s}{2 z} |k_3|^{1/z} k_3^{-1} S(k,\omega + \omega_n) \beta_B  \right] \, \\ 
 &+ \frac{\omega}{4} \sum_n \int \frac{d^2 k d k_3}{(2\pi)^3} \rm{tr} \left[S(k,\omega_n)\frac{s}{2 z}|k_3|^{1/z}k_3^{-1} \beta_B S(k,\omega_n) C^{-1}\beta_A S(k,\omega_n) \right] \, ,
\end{aligned}
\end{equation}
\begin{equation}
\begin{aligned}
G^{\Sigma \Sigma}_{AB}(\omega) &= \frac{1}{\beta}\sum_n \int \frac{d^2 k d k_3}{(2\pi)^3} \rm{tr} \left[S(k,\omega_n)C^{-1}k_A \frac{s}{2z} |k_3|^{1/z}S(k,\omega + \omega_n)C^{-1}k_B \frac{s}{2z} |k_3|^{1/z}\right] \, \\ 
 & +\omega \sum_n \int \frac{d^2 k d k_3}{(2\pi)^3} \rm{tr} \left[S(k,\omega_n)C^{-1}k_B \frac{s}{2z} |k_3|^{1/z} k_3^{-1} S(k,\omega + \omega_n)|k_3|^{1/z} k_3^{-1} \frac{s}{z} \beta_A C^{-1} \right] \, , 
\end{aligned}
\end{equation}
\begin{equation}
C_{AB}(\omega)= \frac{1}{\beta}\sum_n \int \frac{d^2 k d k_3}{(2\pi)^3} \rm{tr} \left[ |k_3|^{1/z-2} \frac{1}{4z^2}\left( \omega_n C^{-1} \beta_A \beta_B + k_A C^{-1} \beta_B \right)S(k,\omega_n) \right]
\end{equation}
\end{widetext}
where we have introduced the Majorana propagator $S(p)= \left[ \beta^A p_A + s  M(p)^{1/2z}C^{-1} \right]^{-1}$.
The form of the contact terms, which require quite a lengthy computation, is justified in Appendix \ref{app:seagull}.
At this point the external $\omega= 2 \pi n T$ is a bosonic Matsubara frequency, the Lorentzian continuation is defined by the substitution $\omega \to i(\omega_L + i\epsilon)$ after the sum over internal frequencies has been performed. The retarded and Euclidean Green's functions are then related by
\begin{equation}
G(\omega, \vec{k})= -i G_E(\omega + i\epsilon,\vec{k}) \, .   
\end{equation}
The Matsubara sum over fermionic frequencies is evaluated using the integral representation of the fermionic sums
 \begin{equation}
 \frac{1}{\beta}\sum_n f(\omega_n) = \frac{1}{2}\int_C \frac{dz}{2\pi i} \tanh\left(\beta z/2\right) f(z) \, .
 \end{equation}
where $C$ is a contour encircling the poles of the hyperbolic tangent. By contour deformation the sum
is expressed as a sum over the residues of the poles of the function $f(z)$. Notice that in the case of Majorana fermions no antiparticles are present, so that the sum over frequencies gives half of the result of that for a Dirac fermion.
Following the analysis of the previous section, furthermore, we expect viscosities to scale as 
\begin{equation}
 \eta^\tau \sim T^{2+z} \, , \ \ \eta^\pi \sim T^{3z} \, , \ \ \eta^\Sigma \sim T^{4-z} \, , \ \ \eta^{\pi\Sigma} \sim T^{2+z} \, , 
\end{equation}
thus we may safely drop all of the vacuum contributions to the thermal sums, since they have no intrinsic parameter which scales under the Lifshitz symmetry.

After this has been done one can divide by the external frequency and safely take the limit of $\omega_L \to 0$. The remaining momentum space integrals are evaluated using the representations
\begin{equation}
\eta_D(s)= \frac{1}{\Gamma(s)} \int_0^\infty dt t^{s-1} n_F(t) \, ,    
\end{equation}
for the Dirichlet eta function and 
\begin{equation}
B(a,b)= \frac{\Gamma(a) \Gamma(b)}{\Gamma(a+b)} = 2 \int_0^{\pi/2} d \phi \sin(\phi)^{2b-1}\cos(\phi)^{2a-1}  \, ,
\end{equation}
for the Euler beta function. We give details of the various computations in Appendix \ref{app:sums}.

The final results read
\begin{align}
\eta^\pi &= \frac{s}{4 \pi^2} \frac{z}{3z+1} T^{3z}  \Gamma(3z) \eta_D(3z) \, , \\
\eta^{\tau} &= \frac{s}{4\pi^2} T^{2+z} \frac{z (z+4)}{(z+1)(z+3)}  \Gamma(z+2)\eta_D(z+2) \, , \\ 
\eta^{\pi \Sigma} &= \frac{s}{4\pi^2} T^{2+z}  \frac{(z+4)}{(z+1)(z+3)}  \Gamma(z+2)\eta_D(z+2) \, , \\
\eta^\Sigma &= \frac{s}{4 z\pi^2} T^{4-z} \frac{(6-z)}{(5-z)(3-z)}  \Gamma(4-z)\eta_D(4-z) \, .
\end{align}
Notice that it holds $\eta^\tau= z \eta^{\pi\Sigma}$. Furthermore, by rescaling $\Sigma \to z \Sigma$ the last three viscosities obey the compact relation
\begin{equation}
\eta^\mathrm{Hall}(\xi)= z \frac{s}{4 \pi^2} T^\xi \frac{(\xi+2)}{(\xi+1)(\xi-1)} \Gamma(\xi)\eta_D(\xi) \, , \label{generalvisco}
\end{equation}
being $\xi$ their Lifshitz scaling dimension.

Notice that all of the coefficients are proportional to the time reversal-breaking parameter $s$, as it should be.

\subsection{A Chern-Simons interpretation as \texorpdfstring{$z\to 0$}{z->0} }

 The values of the viscosities do not seem to bear any universality, since they explicitly depend on the Dirichlet eta function which regulates the thermal sums.
However, at least for the anisotropic momentum current, a suggestive interpretation of the result may be given when the scaling exponent $z$ approaches zero.
In this case the temperature dependence of $\eta^\pi$ vanishes and one may hope to derive an effective action for the result within the framework of effective field theory.
Also if one uses equation \eqref{generalvisco} for the remaining three viscosities, they all vanish in this limit. On the other hand
\begin{equation}
\lim_{z\to 0} \eta^{\pi}  = \frac{s}{24 \pi^2} \, . \label{ztozeropi}
\end{equation}

The intuition behind the $z \to 0$ limit is that the system actually undergoes a dimensional reduction. This can be seen, for example, by computing the density of states with energy for the single particle excitations $\rho(\epsilon) \sim \epsilon^{1+z}$.

In this case the effective action is described by a 2+1 dimensional field theory on a manifold which is obtained by integrating the (possibly non-compact) anisotropic direction.
Indeed \eqref{ztozeropi} is consistent with a Chern-Simons type of action
\begin{equation}
S_{CS} = \kappa \int l \wedge d l \, , \ \ \ \kappa=\frac{s}{48 \pi^2}
\end{equation}
as can be checked by functional differentiation. In this case the Chern-Simons level needs not to be quantized, since the symmetry is associated with is non-compact

There are various way to interpret this phenomenon, and we give two complementary explanations.
They are both based on the idea that our model can be seen as an (infinite) tower of massive Majorana fermions. Notice that as $z \to 0$ the mass $\mu(k_3)= |k_3|^z$ is either arbitrarily small or big depending on whether $k_3 <1$ or $ k_3 >1$. However such masses are all mapped between each other by the Lifshitz symmetry, and should give the same contribution to the effective action.

This is analogue to the fact that, once a massive Dirac fermion is integrated out in $2+1$ dimensions, it generates a Chern-Simons theory with half quantized level $- \rm{sgn} (m) /2$. The half quantization in our case is still present, since the fact that our fermions are Majoranas balances the double occurrences of positive masses (for $k_3 >0$ and $k_3 <0$). As usual, one should extract the value for the Hall coefficients by comparing the result with the one with inverse sign for the mass ($s \to -s$ in our case) and subtract them.

The coefficient $1/48$ is explained as follows. Imagine that the anisotropic direction is compact, in our limit its radius is a number which we may set to one. Fermionic modes on this circle has quantized momenta in half-integer units $k_3= 2n-1$, $n=1,2,...$ . These numbers can be interpreted as the charge of the particle under the translation current $\pi_A$. They enter the Chern-Simons action through their value squared as in the case for $e^2$ in the quantum Hall effect. This gives an infinite sum
\begin{equation}
\kappa= -\frac{s}{4 \pi} \frac{1}{2\pi} \sum_{n=1}^\infty (2n-1)^2 \, , 
\end{equation}  
where the further factor $1/2\pi$ comes from the $dk_3$ integral converted into a sum. The sum over charges may be regulated by zeta function regularization to give $\sum_{n=1}^\infty (2n-1)^2= 4(\zeta(-2)-\zeta(-1)) +\zeta(0)= \frac{1}{3}-\frac{1}{2}=-\frac{1}{6}$ so that
\begin{equation}
\kappa= \frac{s}{48\pi^2} \, ,
\end{equation}
as anticipated.

Another way to find the same result is to use the quadratic form for the Chern-Simons coefficient at finite temperature \cite{Babu:1987rs}:
\begin{equation}
\kappa(\mu(q))= -\frac{1}{8\pi} s \tanh(\beta \mu(q)/2) q^2 \, ,    
\end{equation}
where we take $\mu(q)=q^z$ and take the limit $z \to 0$ afterwards, in this way the temperature acts as a UV regulator.
Integrating over the modes this gives
\begin{equation}
\kappa= -\int_0^\infty \frac{dq}{2\pi} \frac{1}{8\pi} s \tanh(\beta \mu(q)/2) q^2 \, ,
\end{equation}
regulating to zero the vacuum contribution this reduces to 
\begin{equation}
\kappa = T^{3z} \frac{s}{8\pi^2} z \int_0^\infty dt t^{3z-1} n_F(t)= \frac{s}{24\pi^2} \eta(3z) \Gamma(3z+1) T^{3z} \, ,  
\end{equation}
which tends to the previous value as the limit of small $z$ is taken.

One could go one step further and take the Hall conductivity of the single fermion to be quantized as $ n^2 \sigma_H^{2D}$ thus we may conjecture the relationship
\begin{equation}
\eta^{\pi} =  2 \kappa = - \frac{1}{6 \pi} \sigma_H^{2D} + O(z) \, .
\end{equation}
We were not able to find such a nice interpretation for the remaining viscosities.

Why is the Chern-Simons interpretation valid in such limit? A partial explanation comes from the symmetry algebra of our Lifshitz system. In fact, apart from the $ISO(1,2)$ commutation relations, the anisotropic momentum $\Pi= \int \pi_t$ appears only through the nontrivial commutator with the Lifshitz generator $D$ 
\begin{equation}
[D, \Pi]=-z\Pi \, ,
\end{equation} 
and is otherwise a central element. One thus clearly sees that, as $z \to 0$ this commutator vanishes and $\Pi$ behaves effectively as an abelian charge, which is dimensionless.
Since in this limit the system dimensionally reduces to 2D, it is possible that a nonzero Hall conductivity may develop for such abelian charge in the presence of massive fermions, this is the essence of the reason why $\eta^\pi$ does not vanish.

Incidentally, the fact that such a Chern-Simons interpretation may be given tells us that a magnetic torsion $m$ will induce a momentum density $\pi^A v_A \equiv \pi_t$ in the anisotropic direction given by
\begin{equation}
\pi_t = 2 \kappa \  m \, , 
\end{equation}
integrating the above equation relates the total anisotropic momentum with the line integral of the Burgers vector in the anisotropic direction.

Notice also that is an analogue effect to the chiral vortical conductivity for Weyl fermions, once the torsion is rewritten in terms of the ambient metric $g_{\mu\nu}$. For $z \neq 0$ $\kappa$ does not coincide with (twice) the viscosity $\eta^\pi$, as the finite temperature summation is not the same if the limits of zero frequency and momentum are interchanged. 
It can however be easily computed by the same token as before to be
\begin{equation}
\kappa = \frac{s}{8\pi^2} T^{3z} z \Gamma(3z) \eta(3z) = 2 (3z+1) \eta^\pi \, ,
\end{equation}
we will however report on the physics of such effects elsewhere.

\section{Discussion}\label{sec:discussion}
We have shown that quantum critical fermionic Lifshitz fixed point in general possess a non-vanishing Hall viscosity. 
Due to its dissipation less nature it is possible compute these particular transport coefficients at weak coupling. 
Signature of such exotic transport should therefore be measurable even when a an essentially non-interacting
quasiparticle description applies. Signatures of two dimensional Hall viscosity in graphene in a magnetic field have
recently been reported in \cite{berdyugin2018measuring}. It will be interesting to see if the Hall viscosities reported 
here can measured in three dimensional materials along similar lines. 

While the Hall viscosities found here do not seem to bear any universality in general, one of its components may be given a Chern-Simons interpretation in the limit $z \to 0$, in which case we have related it to the intrinsic 2D Hall conductivity of the dimensionally reduced system.

Also, the kind of torsional response we have uncovered is extremely reminiscent of the (much debated) torsional contribution to the mixed anomaly in 3D by the Nieh-Yan term \cite{nieh1982identity,chandia1997topological}
\begin{equation}
 NY[e] = T^a \wedge T^a - R(\omega)_{ab} \wedge e^a \wedge e^b \, ,
\end{equation}  
which in our case should reduce to
\begin{equation}
NY[l]= T \wedge T \, .
\end{equation}
Such contribution has been studied in the context of quantum Hall systems and Weyl semimetals in various occasions \cite{PhysRevLett.107.075502,PhysRevD.88.025040}, however, because of dimensionality reasons, it always comes together with an unspecified UV scale which makes its interpretation very difficult.

In this case, however, the UV scale is represented by the mass $m$, which is a dimensionless quantity from the perspective of the Lifshitz theory. Thus Lifshitz fixed points may provide a more natural setup to relate torsion to the underlying anomalies of the quantum field theory.
In particular, as the exponent $z$ approaches zero, it makes sense (on dimensional grounds) to write an equation like
\begin{equation}
\left( \nabla_\mu -2 G_\mu \right) \pi^\mu = c_\pi \epsilon^{\mu\nu\rho\sigma} T_{\mu\nu} T_{\rho\sigma}\, ,
\end{equation}
since in this limit $l$ does not scale under the Lifshitz symmetry, as an abelian connection should. Consistency of this Ward identity demands $c_\pi = \eta^\pi/8$.

We have studied a broad class of Lifshitz critical point, with arbitrary scaling exponent $z \leq 1 $. We can obtain such models as relevant (in the UV) deformations of the Weyl semimetal model, although subject to an increasing number of fine tuning conditions (see Appendix \ref{app:wsmdetails} for further discussion). It would be interesting to see whether a lattice realization of such low energy theories may also be given.
 
Finally we note that it should be interesting to work out the full Lifshitz hydrodynamics including all the dissipative and possible additional non-dissipative transport coefficients. 

\acknowledgments

This work has been supported by the follwoing grants FPA2015-65480-P(MINECO/FEDER), PIC2016FR6/PICS07480,  and Severo Ochoa Excellence Program grant SEV-2016-0597. The work of C.C. is funded by Fundaci\'on La Caixa under ``La Caixa-Severo Ochoa'' international predoctoral grant. We thank B. Bradlyn, M. Chernodub, A. Cortijo, M.A.H. Vozmediona for discussions.

\section{Appendix}\label{sec:appendix}
\subsection{Details of the four band model}\label{app:wsmdetails}
In this appendix we review some details about the four band model for the WSM-insulator transition, together with some of the salient features of the critical theory.
We start with the four band Lagrangian
\begin{equation}
\mathcal{L}= \bar{\psi} \left( i \gamma_\mu \partial^\mu -m + \gamma_\mu \gamma_5 b^\mu \right) \psi \, , \label{WeylLagrangian}
\end{equation}
Which can be interpreted as a massive Dirac fermion in an axial background $\langle A^5_\mu \rangle=b_\mu$. Standard computations lead to the spectrum of the theory
\begin{equation}
\epsilon(k)^2_{\pm}= k^2 + m^2 +b^2 \pm 2|b| \sqrt{m^2 + (\hat{b}\cdot k)^2} \, . 
\end{equation}
The bands responsible for the low energy behavior are those for which the minus sign above is chosen. The low energy phase is determined by the respective magnitude of $b$ ,$m$. For $|b|>|m|$ the lowest bands touch at $\vec{k}_\pm  = \pm \alpha \vec{b}$, being $\alpha = \sqrt{1- m^2/b^2}$ the screening factor for the chiral charge. In the opposite case the system is gapped, with the gap given by $\Delta_{gap}= 2 \sqrt{m^2-b^2}$.
Since we will mostly an effective description of the two lowest bands, we should notice that the gap between these and the upper one is given by
\begin{equation}
\Delta_{EFT}= \min_{k} \epsilon_+ (k) - \epsilon_-(k) = 2 \max ( |m|, |b|) \, ,
\end{equation}
thus we should always think of our results as valid below these scales. This means, for example, that in the thermal case the temperature should always be much smaller than $ \Delta_{EFT}$.
Of particular interest for us will be the point $|m|=|b|$ at which the lowest bands have the approximate dispersion relation (near $k=0$) 
\begin{equation}
\epsilon^{2}/m^2 = \frac{k_{\perp}^2}{m^2} + \frac{(k \cdot \hat{b})^4}{4 m^4} + O\left((k\cdot b /m)^6\right) \, ,
\end{equation}
which exhibits $z=1/2$ Lifshitz scaling as long as we lie below $\Delta_{EFT}$.
As one can clearly see, the parameter $m$ has no dimensions from the point of view of the Lifshitz scaling and was thus omitted in the main text. However, in order to connect the results presented with the complete field theoretical answer one needs to reintroduce it explicitly. This simply amounts to have all the quantities scale in the right way according to the UV counting, in which $m$ has dimensions of energy.
In particular, the matrix $M^{1/2z}$ in the fermionic Lagrangian gets replaced by $\frac{1}{m^{1/z-1}}M^{1/2z}$. The viscosities also scale with $m$. In this case the trick is to substitute $T$ with the UV dimensionless quantity $\tau= T/m$ and remember that viscosity have dimensions of energy cubed, then
\begin{align}
\eta^\tau &\sim m^{1-z}T^{2+z} \, , \ \ \ \eta^{\pi} \sim m^{3-3z} T^{3z} \, , \\
\eta^{\Sigma} &\sim m^{z-1} T^{4-z} \, , \ \ \ \eta^{\pi \Sigma} \sim m^{1-z}T^{2+z} \, , 
\end{align}
while for our realization of the $z=1/2$ theory we have a definite interpretation for the parameter $m$, for different values of the anisotropic scaling exponent $m$ will in general depend on the particular UV completion one will choose. The appearence of an ultraviolet scale should not be surprising as this is often the case when dealing with torsionful theories. However we stress that from the perspective of the critical point alone, such scale does not play any physical role.

We may furthermore generalize the model \eqref{WeylLagrangian} to support critical points with critical exponent $z= 1/2n$, $n\geq 1 $. The idea is to add couplings to the higher spin counterparts of the chiral current $j_5^\mu$
\begin{equation}
j_5^{\mu_1...\mu_s}= \rm{Str} \left[\bar{\psi}\gamma_5 \gamma^{\mu_1}\overset{\leftrightarrow}{\partial}^{\mu_2}...\overset{\leftrightarrow}{\partial}^{\mu_s} \psi \right] \, ,
\end{equation}
where $\rm{Str}$ refers to the symmetric traceless projection of the tensor. As for the chiral current, such higher spin counterparts are not conserved in the presence of a nonvanishing mass and will in general be irrelevant deformation of the infrared physics.
However let us examine
\begin{equation}
\mathcal{L} = \bar{\psi} \left( i \gamma_\mu \partial^\mu -m\right)\psi + \sum_{s=1} b_{\mu_1...\mu_s} j_5^{\mu_1...\mu_s}  \, ,
\end{equation}
the requirement of maintaing at least $SO(1,2)$ symmetry forces $b_{\mu_1...\mu_s} = b_s b_{\mu_1}...b_{\mu_s}$. The energy dispersion relation then becomes
\begin{equation}
\epsilon(k)^2_{\pm}= k^2 + m^2 +b(k)^2 \pm 2|b(k)|\sqrt{m^2 + (\hat{b}\cdot k)^2} \, ,
\end{equation}
where we have defined $b(x)= \sum_{s=1} b_s (x \cdot \hat{b})^{s-1}$. 
We would like to choose the $b$ function such that a critical point of Lifshitz scaling $z =1/2n$ is reached for small momenta, furthermore, we would like to have to tune only a finite number of current couplings $b_s$ to achieve such result.
We thus put the momentum in the orthoghonal directions to zero and solve the scaling equation ($k_3=k\cdot \hat{b}$)
\begin{equation}
k_3^2 +m^2 + b(k_3)^2 - 2|b(k_3)|\sqrt{m^2 + k_3^2} = k_3^{2n}f^2(k_3) \, ,
\end{equation}
for $b(k_3)$, subject to the requirement that $f^2(k_3)$ is finite at $k_3=0$. This gives, supposing $b, f >0$
\begin{equation}
b(k_3)= \sqrt{m^2 + k_3^2} - k_3^n f(k_3) \, ,
\end{equation}
at this point we may series epand $b(k_3$ and $f(k_3)=\sum_s f_s k_3^s$ and fix the first $2n$ coefficients to match the expression on the rhs. Furthermore, without loss of generality, we may set $b_s=0$ for $s>2n$ and thus fix the function $f$.
The final result is a low energy Lifshitz fixed point with $z=1/n$, obtained by tuning $2n$ parameters through
\begin{align}
b_{2s}&=m^2 \frac{(1/2)_s}{s!} \left(\frac{k_3}{m}\right)^{2s} \, \ \ s\leq n \, \\
f_{2s}&= m^2 \frac{(1/2)_{s+n}}{(s+n)!} \left(\frac{k_3}{m}\right)^{2s} \, .
\end{align}
Of course such a critical point still has a hige amount of fine tuning. Furthermore, it can be continuously reached by deforming the $z=1/2$ critical point withouth breaking any further symmetries. In this sense we expect the physics at different $z$ to belong to the same universality class.

It could be interesting to see whether less fine tuned versions of such critical points exist, and if so what is their lattice realization.
 
\subsection{Seagull terms}\label{app:seagull}
Before moving to the calculation itself, it is however important to verify whether any contact (Seagull) term may arise from the dependence of the strain tensor on connection and torsion.
Seagull terms typically arise in quantum field theory due to the explicit dependence of the curved spacetime stress tensor and currents on the spin or Christoffel connection.
This causes functional differentiation to give rise to terms proportional to
\begin{equation}
\delta^\rho_\mu \delta_D^A \frac{\delta}{\delta e_\rho^D(x)} {\omega_\nu}^{BC}(y)\equiv Z_{\mu\nu\alpha}^{ABC} \partial^\alpha \delta(x-y) \, ,
\end{equation}
where, in the flat spacetime limit,
\begin{equation}
Z_{\mu\nu\alpha}^{ABC}= \frac{1}{2}\left( \eta_{\nu\alpha} \delta^{B}_\mu \delta_A^C + \delta_\alpha^C \delta^B_A \delta_\nu^\mu - \delta_\alpha^B \delta^C_\mu \delta_\nu^A  \right)  - (B \leftrightarrow C) \, .  
\end{equation}
These contribute to the linear response theory with finite terms, that are computed from a one-loop diagram with no external momenta present. In particular the external momentum is carried by the derivative of the delta function, so that in order to compute viscosities we set $\alpha=0$.
Apart from these, other contact terms may arise by functional differentiation of the vielbein itself. We will disregard such contributions.

Let us start from the correlators of two $\tau$. In this case one has to compute the classic contact term of a free fermionic stress tensor. This is a well known computation, see for example \cite{Manes:2012hf}, the final result gives:
\begin{equation}
C_{ABCD}(x,y) = -\frac{i}{16} \delta_{AB} \chi^T(x) \left( \left\{ \frac{1}{4}[\beta_B, \beta_D] , \beta_0 \right\}\right) \chi(x) \partial_0 \delta(x-y) + A \leftrightarrow B \, , C \leftrightarrow D
\end{equation}
which in momentum space gives the contact term integral we will compute in the next section.  
There are three further cases to be examined. The first is the correlators of two anisotropic momentum currents $\pi_A$, $\pi_B$. 
Seagull terms in this case arise from the dependence of the anisotropic current on torsion. Since we work with the $SO(1,2)$ connection only, no such dependence arises in the covariant derivative and the contact term vanishes.

A second contact term may contribute to the $\Sigma_A \Sigma_B $ correlator due to the vielbein dependence of $\Sigma$. To start, recall that in position space this reads
\begin{equation}
C^{AB}= \frac{\partial \Sigma_A}{\partial {\omega_\nu}^{CD}} Z_{\mu\nu\alpha}^{BCD} \partial^\alpha \delta(x-y) l^\mu \, ,
\end{equation}
where the last $l^\mu$ projects on the right component of the vielbein variation. We will be interested of the part of said contact term which is proportional to $\epsilon_{AB}$, thus enconding the nondissipative viscosity.
First one may notice, using the expression above for $Z$, that
\begin{equation}
Z_{\mu\nu\alpha}^{BCD} l^\mu = \frac{1}{2} l_\nu \delta_\alpha^D \delta^{BC} - (B \leftrightarrow C) \, , \label{ZetaSigma}
\end{equation}
thus the only contributions to the contact term will come from derivatives $\nabla_l$ in $\Sigma$. The contributions may be divided in two parts, the first stemming from the unimproved strain $ \hat{\Sigma}$ and the latter from the improvement term coming from the spin current.

For the first term we have, using \eqref{unimprovedstrain}
\begin{equation}
\begin{aligned}
\frac{\partial \hat{\Sigma}_A }{\partial \omega_\nu^{CD}} &=\frac{s}{z}\left(\frac{1}{2z} -1\right)l^\nu \chi^T \left[ \overleftarrow{\partial}_l \overset{\leftrightarrow}{\partial}_{A} C^{-1} \beta_C C^{-1} \beta_D C^{-1} +\overrightarrow{\partial}_l \overset{\leftrightarrow}{\partial}_{A} \beta_C C^{-1} \beta_D   \right] M^{1/2z -2}\chi \\
& + \frac{s}{z}\chi^T \left[\overleftarrow{\partial}_A  C^{-1} \beta_C C^{-1} \beta_D C^{-1} + \overrightarrow{\partial}_A \beta_C C^{-1} \beta_D  \right]M^{1/2z-1}\chi
\end{aligned}
\end{equation}
up to terms orthogonal to $l^\nu$. Going to momentum space and remembring that one of the two $\beta$ matrices is the identity because of \eqref{ZetaSigma}, one is left with an anticommutator $\beta_D C^{-1} + C^{-1} \beta_D=0$ if $D$ is spatial. Si the whole contribution vanishes.
Thus the possible contact terms may come from the improvement only. 

The second term gives
\begin{equation}
\begin{aligned}
\frac{\partial \Sigma_{\rm imp}^A }{\partial \omega_\nu^{CD}} &= \frac{s}{z}l^\nu \partial_B \chi^T\left[\left(\frac{1}{2z}-1\right)M^{1/2z-2}\overset{\leftrightarrow}{\partial}_l \left(\overleftarrow{\partial}_l \gamma^{BA} C^{-1} \gamma^{CD} - \overrightarrow{\partial}_l \gamma^{CD} \gamma^{BA} C^{-1} \right)\right]\chi \\
&+ \frac{s}{z} l^\nu \partial_B \chi^T\left[ M^{1/2z-1} \left(\gamma^{BA} C^{-1} \gamma^{CD} + \gamma^{CD} \gamma^{BA} C^{-1}\right)   \right] \chi \, .
\end{aligned}
\end{equation}
This simplifies in a considerable way in momentum space, where the two contributions above sum if the external frequency is set to zero. The result is
\begin{equation}
\frac{\partial \Sigma_{\rm imp}^A }{\partial \omega_\nu^{CD}}(q)= l^\nu \frac{s}{2 z^2} q^B \chi^T |q \cdot l|^{1/z-2} X^{CD}_{AB} \chi \, , 
\end{equation}
with
\begin{equation}
X^{CD}_{AB}= \gamma^{BA} C^{-1} \gamma^{CD} + \gamma^{CD} \gamma^{BA} C^{-1} \, .
\end{equation}
the expression for $X$ can be recasted as either a commutator or an anticommutator depending on whether $CD=0i$ or $CD=ij$.
In our case the relevant part will be
\begin{equation}
X^{CD}_{AB}= [\gamma^{CD},\gamma_{AB}]C^{-1}\left(\delta^C_0 - \delta^D_0 \right) \, ,
\end{equation}
one may now use the Lorentz algebra
\begin{equation}
[\gamma_{CD},\gamma_{AB}]= \eta_{CA}\gamma_{DB} + ({\rm cyclic}) \, , 
\end{equation}
to simplify the expression further. The final result taking into account the fact the either $C$ or $D$ are in the time direction, reads
\begin{equation}
\frac{\partial \Sigma_{\rm imp}^A }{\partial \omega_\nu^{CD}}(q)=l^\nu \frac{s}{4 z^2} \chi^T |q \cdot l|^{1/z-2}\left(q_0 C^{-1} \beta^ A \beta_D + q_D C^{-1} \beta^A  \right) \delta^C_0  \chi \, .
\end{equation}

We will use this term in the next section for the computation of the linear response.

One last contact term may come from the $\Sigma_A$ $\pi_B$ correlator, and can be seen either through the torsion dependence of $\Sigma_A$ or through the spin connection dependence of $\pi_A$.
The first of the two is simpler to compute, in this case, in fact, the only dependence on torsion comes from the $G^B \sigma_{BA}$ term in the definition of $\Sigma$, recalling the definition of $G_\mu$ one has
\begin{equation}
\delta G_\mu = -l^\nu \left(\partial_\nu \delta l_\mu - \partial_\mu \delta l_\nu \right) -\delta l^\nu (dl)_{\nu\mu} \, ,    
\end{equation}
this gives a seagull contribution to $\eta^{\pi\Sigma}$ only if the derivative is in the time direction and $\delta l_\mu$ is in a spatial direction. This is however not possible, since the only time derivative comes with the anisotropic component of $l_\mu$ which does not contribute to the correlator we are interested in.

\subsection{Relevant Feynman graphs and Matsubara sums}\label{app:sums}
In this section of the supplemental material we review the detailed calculations of the 3D Hall viscosity. The main steps of the procedure have already been outlined in the main text in section \ref{sec:Hviscosity}.
Here we reproduce the essential details of the computations

\subsubsection{Computation of \texorpdfstring{$\eta^\pi$}{etapi}}
We start with the computation of the $\pi_A \pi_B$ correlator. Since we are interested only in the contributions to the Hall viscosity tensor we will always implicitly extract the part of the correlators that goes like the appropriate projector.
The $\pi_A \pi_B$ correlator in computed by the Lorentzian continuation of the following Euclidean diagram
\begin{equation}
\langle \pi_A(-\omega,0) \pi_B( \omega,0) \rangle=\frac{1}{\beta} \sum_n \int \frac{d^2 k d k_3}{(2\pi)^3} \rm{tr} \left[S(k, \omega_n) \beta_A S(k,\omega + \omega_n) \beta_B k_3^2 \right] \label{pipicorr}
\end{equation}
where $\omega= 2\pi m T $ is a bosonic Matsubara frequency, while the discrete sum runs over fermionic frequencies $\omega_n = (2n+1)\pi T$. In Majorana notation the fermionic propagator is
\begin{equation}
S(p)=\left( \beta^A p_A + s M(p)^{1/2z}C^{-1} \right)^{-1} \, , \beta^A = C^{-1} \gamma^A \, .
\end{equation}
 Due to the Majorana nature of the computation and thus the absence of antiparticles, the Matsubara sums will only give half of the expected result, as it can be explicitly checked that the poles for particles and antiparticles give the same contributions to the odd viscosity.

To begin we have to evaluate the trace over the Dirac indeces to extract the odd projector.
We will often encounter such traces in the various computations, in this case the key result is that 
\begin{equation}
\rm{tr}\left[ \beta_A \beta_B C^{-1}\right]= 2 \epsilon_{AB} \, ,
\end{equation} 
 where $\epsilon_{AB}=\epsilon_{ABC}u^C$ and $u^C$ represents the time direction. This can be readily checked via the representation $\beta_0= -1$, $\beta_1= -\sigma_x$, $\beta_2 = \sigma_z$, $C=- i \sigma_y$, which we will use in practical computations.

In \eqref{pipicorr} one readily sees that the trace can be saturated only in the case in which we have an $M(k)$ contribution from the first propagator and a $\beta_C \omega^C \equiv - \omega$ one from the second. The contribution from the internal Matsubara frequency cancel because of the ordering of the matrices. 
 
 The Hall contribution then reads
 \begin{equation}
 \langle \pi_A(-\omega,0) \pi_B( \omega,0) \rangle_H = \epsilon_{AB} \omega  \frac{4 s}{4 \pi^2} \int_0^{\infty} d k_3 k_3^{1/z+2} \int_0^{\infty} dk k  g(\epsilon, \omega) \, , \label{pipicorr2}
 \end{equation}
 where
 \begin{equation}
g(\epsilon,\omega) = \sum_n \frac{1}{\omega_n^2 +\epsilon^2(k,k_3)} \frac{1}{(\omega+\omega_n)^2 + \epsilon^2(k,k_3)} \, , \ \ \epsilon^2(k,k_3)= k^2 + k_3^{2/z} \, ,
 \end{equation}
 is the Matsubara sum.
Its evaluation of the Matsubara sum is straightforward and gives 
 \begin{equation}
g(\epsilon, \omega)= - \frac{\tanh(\beta \epsilon/2)}{8 \epsilon (\epsilon^2 + \omega^2/4)} \, ,
\end{equation} 
where we stress that $\omega$ must be kept as a bosonic Matsubara frequency.
 
 We'll be eventially interested in continuing the result to the Lorentzian sector to extract the retarded propagator. This is done as costumary by the replacement $\omega = 2 \pi m T \to i (\omega_L +i 0)$, followed by the $\omega_L \to 0 $ limit.
 However in this case, since the transport we are interested is nondissipative, we expect the density of states $\rho_{AB}^{\pi\pi}(\omega)= \rm{Im} G^{\pi\pi}_{AB}(\omega)$ to vanish as the frequency is set to zero.
 This can be explicitly checked by computing the residue of the integrand of $G^{\pi\pi}$, which scales as $\omega^{3z +1}$, so that both its value and its derivative's vanish in the zero frequency limit. A similar reasoning hold for the other integrals.
We may then take the naive $\omega \to 0$ limit inside the integral after performing the Matsubara sums.
 
At this point we divide vacuum from thermal contributions through the identity
\begin{equation}
\tanh(x/2)= 1- 2 n_F(x) \, , 
\end{equation}
where $n_F(x)= 1/(1+ e^{x})$ is the Fermi-Dirac distribution. Since the vacuum has no intrinsic Lifshitz scaling parameter, its contribution vanishes in any sensible regulation scheme.
On the other hand, the thermal part gives the Hall conductivity to be
\begin{equation}
\eta^\pi = \frac{s}{4\pi^2}  \int_0^\infty d k_3 k_3^{1/z} \int_0^{\infty} dk k \frac{n_F(\beta \epsilon(k,k_3))}{\epsilon(k,k_3)^3} \, .
\end{equation}
We now change variables to $u= \beta k_3^{1/z}$, $v=\beta k$ to get
\begin{equation}
\eta^\pi = \frac{s}{4\pi^2} T^{3z} I_{3z} \, ,
\end{equation}
where
\begin{equation}
\begin{aligned}
I_{3z}&= z \int_0^\infty du u^{3z} \int_0^\infty dv v \frac{n_F( \sqrt{u^2 + v^2})}{\left(u^2+ v^2 \right)^{3/2}}= \\
&= z \int_0^\infty d\rho \rho^{3z-1} n_F(\rho) \int_0^{\pi/2} d\phi \sin(\phi) \cos(\phi)^{3z} = \frac{z}{3z+1}\Gamma(3z) \eta_D(3z) \, ,
\end{aligned}
\end{equation}
by going to polar coordinates $u=\rho \cos(\phi)$, $v=\rho \sin(\phi)$. Finally
\begin{equation}
\eta^\pi = \frac{s}{4\pi^2} T^{3z} \frac{z}{3z+1}\Gamma(3z) \eta_D(3z) \, .
\end{equation}

Most of the other computations go along the same lines, in particular we will make the same series of changes of variables, as well as computing largely the same Matsubara sums. We will thus focus on the technical differences to speed up the presentation.

\subsubsection{Computation of \texorpdfstring{$\eta^{\pi\Sigma}$}{etasigma}}

We proceed to compute the Hall conductivity stemming from the correlator between $\pi$ and $\Sigma$. In this case the contribution splits into two parts, the first one given by the unimproved $\Sigma$, $\hat{\Sigma}_A= \frac{s}{z }\chi^T M^{1/2z -1} \left( \overleftarrow{\partial}_\nu l^\nu \overrightarrow{\partial}_A +\overleftarrow{\partial}_A  l^\nu \overrightarrow{\partial}_\nu \right) C^{-1} \chi $ and a second one coming from the improvement term $\partial^B\sigma_{BA}$.
The first of the two is given by the graph
\begin{equation}
\langle \hat{\Sigma}_A(-\omega,0) \pi_B( \omega,0)\rangle = \frac{1}{\beta}\sum_n \int \frac{d^2 k d k_3}{(2\pi)^3} \rm{tr} \left[S(k, \omega_n)C^{-1} k_A \frac{s}{ z } |k_3|^{1/z-2} k_3 S(k,\omega + \omega_n) \beta_B k_3 \right] \, .
\end{equation}
The trace is evaluated in a similar way as before, only that now we will need a $\beta_C \omega^C$ and a $\beta_D k^D$ contribution from the propagators. The trace will be proportional to $-2\epsilon_{BD}k^D \omega$. Performing the angular integral $d^2 k$ amounts to the substitution $k_A k^D \to \delta_{A}^D k^2$ and a factor of $2\pi$, so
\begin{equation}
\langle \hat{\Sigma}_A(-\omega,0) \pi_B( \omega,0)\rangle = \frac{2 s}{4\pi^2} \epsilon_{AB} \omega  \int_0^{\infty} d k_3 k_3^{1/z} \int_0^\infty dk k^3 g(\epsilon,\omega) \, ,
\end{equation}
using the previous change of variables this gvies
\begin{equation}
\eta^{\pi\Sigma}(2pf)= \frac{2 s}{4 z \pi^2}T^{2+z} I_{z+2}\, ,
\end{equation}
where
\begin{equation}
\begin{aligned}
I_{2+z}&= \frac{z}{4} \int_0^{\infty} du u^{z} \int_0^{\infty} dv v^3 \frac{n_F(\sqrt{u^2 +v^2})}{\left( u^2 +v^2 \right)^{3/2}} = \\
 &= \frac{z}{4} \int_0^\infty d \rho \rho^{z+1} n_F{\rho} \int_0^{\pi/2} d \phi \sin(\phi)^3 \cos(\phi)^{z}  =\frac{1}{2(z+1)(z+3)}\Gamma(z+2)\eta_D(z+2) \, ,
\end{aligned}
\end{equation}
so
\begin{equation}
\eta^{\pi\Sigma}(2pf)= \frac{ s}{4\pi^2} \tau^{z+2}m^3 \frac{1}{(z+1)(z+3)}\Gamma(z+2)\eta_D(z+2).
\end{equation}
For the improvement term we instead get
\begin{equation}
\frac{1}{2}\langle \sigma_{0A}(-\omega)\pi_B(\omega)\rangle = \frac{1}{4}\sum_n \int \frac{d^2 k d k_3}{(2\pi)^3} \frac{s |k_3|^{1/z}}{z} \rm{tr} \left[S(k,\omega_n)\beta_A C^{-1} S(k,\omega + \omega_n) \beta_B\right] \, ,    
\end{equation}

as both $A$ and $B$ are spatial, the only way to get an $\epsilon$ tensor is that the matrices from the two propagators contract between each other. The trace thus gives
\begin{equation}
\rm{tr} \left[S(k,\omega_n)\beta_A C^{-1} S(k,\omega + \omega_n) \beta_B\right]= 2 \epsilon_{AB}\frac{\omega_n(\omega_n+\omega) + \epsilon(k,k_3)^2}{(\omega_n^2 +\epsilon(k,k_3)^2((\omega_n+\omega)^2+\epsilon(k,k_3)^2}  \, ,  
\end{equation}
which may be simplified, writing $\omega_n(\omega_n+\omega)=1/2(\omega_n^2 +(\omega+\omega_n)^2 -\omega^2)$ to
\begin{equation}
\epsilon_{AB}\left[ \frac{1}{\omega_n^2 +\epsilon(k,k_3)^2} + \frac{1}{(\omega_n+\omega)^2+\epsilon(k,k_3)^2} - \frac{\omega^2}{(\omega_n^2 + \epsilon(k,k_3)^2)((\omega_n+\omega)^2+\epsilon(k,k_3)^2)} \right]  \, ,  
\end{equation}
the first two sums are easily computed $\frac{1}{\beta}\sum_n \frac{1}{(\omega_n + \omega)^2 + \epsilon(k,k_3)^2} = -\frac{\tanh{\beta \epsilon(k,k_3)/2}}{4\epsilon(k,k_3) }$ to be equivalent while the third vanishes in the $\omega \to 0$ limit.
We then get  
\begin{equation}
\begin{aligned}
&\epsilon^{AB}\frac{1}{2}\langle \sigma_{0A}(-0)\pi_B(0)\rangle = -\int \frac{d^2 k d k_3}{4(2\pi)^3}    \frac{s |k_3|^{1/z}}{z} \frac{1}{\epsilon(k,k_3)}\tanh(\beta \epsilon(k,k_3)/2)  \\
&= \frac{s}{4 \pi^2} T^{2+z} \int_0^\infty d u u^z \int_0^\infty dv v \frac{n_F(\sqrt{u^2+v^2})}{\left( u^2 + v^2\right)^{1/2}} = \\
&= \frac{s}{4 \pi^2} T^{2+z} \int_0^\infty d\rho \rho^{z+1} \int_0^{\pi/2} d\phi \sin(\phi)  \cos(\phi)^{z}= \frac{s}{4 \pi^2} T^{2+z} \frac{1}{(z+1)}\Gamma(z+2)\eta_D(z+2) \, , \label{pisigmasecond}
\end{aligned}
\end{equation}
summing the two contributions we finally get
\begin{equation}
\eta^{\pi\Sigma}= \frac{s}{4 \pi^2} T^{2+z} \frac{(z+4)}{(z+1)(z+3)}\Gamma(z+2)\eta_D(z+2) \, .
\end{equation}

\subsubsection{Computation of \texorpdfstring{$\eta^\tau$}{etatau}}
We next move move to the intrinsic $2+1$ dimensional thermal Hall viscosity, for which one should compute both the two point function $\tau \tau$ and the seagull term $C_{ABCD}$.
The first of these is given by the integral
\begin{equation}
\langle \tau_{AB}(-\omega,0) \tau_{CD}( \omega,0) \rangle  =\frac{1}{\beta} \sum_n \int \frac{d^2 k dk_3}{(2\pi)^3} \rm{tr} \left[S(k, \omega_n) \beta_{(A} S(k,\omega + \omega_n) \beta_{(C} k_{B)} k_{D)} \right] \, .
\end{equation}
We are interested in the contribution proportional to $P_{ABCD}$ of this correlator. To get the right factors it is sufficient to work with one combination of indeces, the full structure of the projector is then automatically recovered through symmetrization.
The trace is computed in the same way as for $\eta^\pi$ and we find
\begin{equation}
\langle \tau_{AB}(-\omega,0) \tau_{CD}( \omega,0) \rangle_H = \omega P_{ABCD}  \frac{2 s}{4\pi^2}  \int_0^\infty d k_3 k^{1/z} \int_0^\infty dk k^3 g(\epsilon,\omega)\, =  z P_{ABCD} \eta^{\pi\Sigma}(\rm{2pf}) \, ,
\end{equation}
Confronting this expression with the computation of $\eta^{\pi\Sigma}$ we deduce that the two point function contribution to this component of the visocsity will be given by
\begin{equation}
\eta^\tau(\rm{2pf})=  z \eta^{\pi \Sigma}(\rm{2pf}) = \frac{s}{4\pi^2} T^{z+2}\frac{z}{(z+1)(z+3)}\Gamma(z+2)\eta_D(z+2)  \, .
\end{equation}
to get the full result we still have to evaluate the contact term $C_{ABCD}$.
In momentum space the seagull term gives the following diagram
\begin{equation}
C_{ABCD}(\omega) = \frac{\delta_{AC}}{16}\frac{1}{\beta}\sum_n \int \frac{d^2 k dk_3}{(2\pi)^3} \rm{tr} \left[\beta_{[A} C^{-1} \beta_{B]} \omega S(k, \omega_n)\right] + A \leftrightarrow B \, , C \leftrightarrow D \, ,
\end{equation}
the trace is computed as before and the index structure organizes to give a projector, so
\begin{equation}
\begin{aligned}
C_{ABCD}(\omega) &= \omega P_{ABCD} \frac{s}{4\pi^2} T^{2+z}  z \int_0^\infty d u u^z \int_0^\infty dv v \frac{n_F(\sqrt{u^2+v^2})}{\left(u^2 + v^2\right)^{1/2}} \\
&= \omega P_{ABCD} \frac{s}{4 \pi^2} T^{2+z} \frac{1}{(z+1)}\Gamma(z+2)\eta(z+2)   \, ,
\end{aligned}
\end{equation}
summing all up we get the relation
\begin{equation}
\eta^\tau= z \eta^{\pi\Sigma}= \frac{s}{4 \pi^2} T^{2+z} \frac{z(z+4)}{(z+1)(z+3)}\Gamma(z+2)\eta_D(z+2) \, .  \end{equation}

\subsubsection{Computation of \texorpdfstring{$\eta^\Sigma$}{etasigma}}
Finally we inspect the value of $\eta^\Sigma$, this is the longest computation but we may use most of the tricks learned before to speed it up. It can be divided in three parts: the first coming from the correlators of the unimproved strains $\hat{\Sigma}$, the second coming from the correlator of one of these with the improvement term and the last one stemming from the contact terms. It is simple to convince oneself that the unimproved correlator vanishes.  This is because the Feynman diagram contains a term $k_A k_B$ which should be antisymmetrized. 

The improvement term, on the other hand, behaves in much the same may as we have seen in the $\pi \Sigma$ correlator and gives a contribution
\begin{equation}
\eta^\Sigma_{\rm imp} = \lim_{\omega \to 0} \langle \sigma_{0A}(-\omega) \hat{\Sigma}_B(\omega) \rangle \epsilon^{AB}
\end{equation}
which reads in terms on Feynman diagrams
\begin{equation}
\eta^\Sigma_{\rm imp}= \lim_{\omega \to 0}\frac{1}{\beta} \sum_n \int \frac{d^2 k d k_3}{(2\pi)^3} \rm{tr} \left[|k_3|^{1/z} k_3^{-1} k_B C^{-1} S(k,\omega_n)|k_3|^{1/z} k_3^{-1} \frac{1}{2} \beta_A C^{-1} S(k, \omega_n +\omega) \right] \, ,
\end{equation}
as before, the odd part of the trace may be computed by bringing up one term with the anisotropic momentum and one $\beta$ matrix. This gives
\begin{equation}
\begin{aligned}
\eta^\Sigma_{\rm imp} &= \lim_{\omega \to 0}\frac{1}{z^2}\frac{s}{2\pi^2} \frac{1}{\beta}\sum_n \int_0^\infty d k_3 \int_0^\infty dk k^3 k_3^{3/z-2} g(\epsilon,\omega) \\
&= \frac{s}{8\pi^2 z}T^{4-z} \int d\rho \rho^{3-z} n_F(\rho) \int_0^{\pi/2} d\phi \sin(\phi)^3 \cos(\phi) \\
&= \frac{s}{4 z \pi^2} T^{4-z} \frac{\Gamma(4-z)\eta_D(4-z)}{(5-z)(3-z)} \, .
\end{aligned}
\end{equation}
Finally one has to take care of the contact term, whose form we had computed in the previous section.
In this case one has the Feynman graph
\begin{equation}
\eta^\Sigma_{\rm{ct}}=\frac{1}{2z^2} \frac{1}{\beta}\sum_n \int \frac{d^2 k d k_3}{(2\pi)^3} \rm{tr} \left[ |k_3|^{1/z-2} \frac{1}{2}\left( \omega_n C^{-1} \beta_A \beta_B + k_A C^{-1} \beta_B \right)S(k,\omega_n) \right] \epsilon^{AB}
\end{equation}
the trace gives
\begin{equation}
\rm{tr} \left[ |k_3|^{1/z-2} \frac{1}{2}\left( \omega_n C^{-1} \beta_A \beta_B + k_B C^{-1} \beta_A \right)S(k,\omega_n) \right] \epsilon^{AB}= 1 - \frac{|k_3|^{2/z}}{\omega_n^2 +\epsilon(k,k_3)^2} \, ,
\end{equation}
the first term is a vacuum contribution which may be regulated away, while the second Matsubara sum can be easily computed.
One gets
\begin{equation}
\eta^\Sigma_{\rm{ct}}= \frac{s}{4 z\pi^2} \int_0^\infty d \rho \rho^{3-z} n_F(\rho) \int_0^{\pi/2} d\phi \sin(\phi) \cos(\phi)^{2-z} = \frac{s}{4z \pi^2} T^{4-z} \frac{\Gamma(4-z) \eta_D(4-z)}{(3-z)} \, .
\end{equation}
Putting everything together we finally find
\begin{equation}
\eta^\Sigma= \frac{s}{4z \pi^2} T^{4-z} \frac{(6-z)}{(5-z)(3-z)}\Gamma(4-z)\eta_D(4-z) \, .
\end{equation}
It is nice to notice that the three viscosities $\eta^\tau$, $\eta^\Sigma$ and $\eta^{\pi\Sigma}$ can be compactly re-expressed (provided we renormalize $\Sigma \to z \Sigma$) as functions of their scaling dimension $\xi$  
\begin{equation}
\eta(\xi)/z = \frac{s}{4\pi^2} T^\xi \frac{(\xi+2)}{(\xi+1)(\xi-1)}\Gamma(\xi)\eta_D(\xi) \, .
\end{equation}

\bibliography{AnomTrans}{}

\end{document}